\documentclass[a4paper,11pt]{article}
\usepackage{amsmath}
\pdfoutput=1
\usepackage{jcappub}
\usepackage{ihmacros}
\usepackage{subfig}
\usepackage{booktabs}

\renewcommand{\Pr}{\ensuremath{\mathcal{P}_{\mathcal{R}}}}

\title{Second Order Perturbations During Inflation Beyond
  Slow-roll} 
\author[a]{Ian Huston,} 
\author[a]{Karim A. Malik}
\affiliation[a]{ Queen Mary University of London,\\ Astronomy Unit, School of
  Physics and Astronomy, Mile End Road, London E1 4NS}
\emailAdd{i.huston@qmul.ac.uk} \emailAdd{k.malik@qmul.ac.uk}

\abstract{We numerically calculate the evolution of second order cosmological
perturbations for an inflationary scalar field without resorting to
the slow-roll approximation or assuming large scales. In contrast to
previous approaches we therefore use the full non-slow-roll source term
for the second order Klein-Gordon equation which is valid on all
scales. The numerical results are consistent with the ones obtained previously where
slow-roll is a good approximation. We investigate the effect of
localised features in the scalar field potential which break slow-roll
for some portion of the evolution. The numerical package solving the
second order Klein-Gordon equation has been released under an open
source license and is available for download.  }


\toccontinuoustrue

\newcommand{\figsize}{small}

\newcommand{\twofigwidth}{0.46\textwidth}

\begin{document}

\maketitle

\section{Introduction}

The arrival of more and better data in the last couple of years has
fundamentally changed theoretical cosmology. Until only a few years
ago it was sufficient to use linear order theory to analyse the data
and calculate first order observables, such as the power spectrum;
higher order theory wasn't required.  The new data, in particular the
Cosmic Microwave Background (CMB) anisotropy maps as provided by the {\sc Wmap}
and the {\sc Planck} satellites \cite{wmapwebsite, planckwebsite},
make it possible to go beyond linear order and derive higher order
observables such as the bispectrum from the data. Considerable effort
has therefore been spent on extracting observable signatures beyond
the power spectrum and to extend cosmological perturbation theory
beyond linear order.

So far calculations of higher order observables such as the
bispectrum, or its popular parametrisation $\fnl$, have relied heavily
on approximations. This meant usually assuming large scales, that is
large compared to the ``horizon'', and imposing slow-roll
\cite{book:liddle}. However, these restrictions not only constrain
the validity of the results, but also limit the number of models that
can be studied, excluding many interesting cases.  For example, even in 
linear theory it is already well known that interesting and observable
effects occur when slow-roll is violated at the end of inflation
\cite{Leach:2001zf}.

%%%%%%%%%%%%%%%%%%%%%%%%%%%%%%%%%%%%%%%%%%%%%%%%%%%%%%%%%%%%%%%%

Second order perturbations play a crucial role in our quest to
understand the non-linear physics of the early universe. Previous
works by us and collaborators \cite{Malik:2005cy,Malik:2007nd} have
derived the equations of motion for second order perturbations in the
long wavelength approximation, the slow-roll approximation and also
the full non-slow-roll case for all scales \cite{Malik:2006ir}.

Previously in \Rref{hustonmalik} we numerically solved the
evolution of second order perturbations under the assumption that the
source term, which consists of convolutions of first order
perturbations and their derivatives, is calculated in the slow-roll
approximation. We showed that the source term closely follows the form
of the first order power spectrum. In this paper we go beyond the slow
roll approximation to calculate the full source term equations for a 
single scalar field. This
full treatment is needed in cases where slow-roll is broken. The
results of the updated calculation are consistent with those of our previous work
for the $\msqphisq$ potential for which the slow-roll approximation is
an exceptionally good one. To go beyond slow-roll in this paper, we
study ``step'' and ``bump'' potentials, for which slow-roll is broken as $|\eta_V|>1$ around
the feature ($\eta_V$ denoting one of the slow-roll parameters). 
Step potentials have been widely used to create features in the power spectrum
of inflationary perturbations which might more accurately match the observed power spectrum from
{\sc WMAP}
\cite{Adams:2001vc,Stewart:2001cd,Choe:2004zg,Joy:2008qd,Hamann:2009bz,Hazra:2010ve}.
In this paper we will follow the potential forms described by Chen \etal
\cite{Chen:2008wn,Chen:2006xjb}. In addition to
being able to calculate the second order field perturbation in this
case, the result of the full equations seems to be smoother at very
early times when the perturbations are highly oscillatory (due to a
larger damping term, which is lost in slow-roll).

The final goal of this continuing work is a numerical calculation of
the curvature perturbation at second order for all length scales. This
will probe effects both inside and outside the horizon in a way that
is not possible using other methods, for example the $\delta N$
formalism
\cite{Starobinsky:1982ee,Starobinsky:1986fxa,Salopek:1990jq,Sasaki:1995aw,Lyth:2004gb},
which a priori uses large scales only. The applications of such a
calculation are many and varied and range from an investigation of the
non-Gaussian nature of inflationary perturbations 
to the exploration of higher order effects such as the sourcing of
gravitational waves at second order \cite{Ananda:2006af} and
vorticity generation \cite{Christopherson:2009bt}.

In the current paper we have only considered the effect of a single scalar field. One goal of
future work is to extend the numerical code to deal with more than one field. In a multifield
system the curvature perturbation is not the only
important observable with isocurvature components also playing a
significant role.  Calculating the field evolution is the only way to
gain access to these isocurvature results, in a way that would not be
possible if we had concentrated purely on the curvature
perturbations. With this goal in mind we have fashioned the
numerical calculation in terms of a scalar field instead of the
curvature perturbation which might otherwise have been considered a more observationally relevant
quantity in the single field case.
We have developed the numerical code considerably from that used in
\Rref{hustonmalik}. The change from the slow-roll to full equations
adds significant complexity to the calculation but this has been
countered by optimising the time sensitive functions and implementing
more of them in a compiled language. The numerical code has been
released under an open source license (Modified BSD license) and is
available for download \cite{pyflation}.\\

In Section~\ref{sec:perts} we present the Klein-Gordon equation for
second order cosmological perturbations. This is the central part of
the code described in Section~\ref{sec:code}. The results we have
obtained are in Section~\ref{sec:results} and discussion of these and
future goals is contained in Section~\ref{sec:disc}. Throughout the
text we use $\kapsq$ and label conformal time derivatives with a
prime. We assume a flat Friedmann-Robertson-Walker background
throughout this paper.

\section{Second Order Perturbations}
\label{sec:perts}

\subsection{Klein Gordon Equations}
Cosmological perturbations beyond linear order have been studied in
depth in recent times and a brief summary can be found in
Appendix~\ref{sec:apxsoperts} or at length in
Refs.~\cite{Malik:2008yp,Malik:2008im}  (see Ref.~\cite{Noh:2004bc} for an example of earlier work
in this area). In particular the Klein-Gordon
equations for the background, first order and second order
perturbations are given in
Eqs.~\eqref{eq:KGback-num}-\eqref{eq:KG2flatsingle-num}. These
equations are given in real space, however, and in general the
dynamics of the scalar field become clearer in Fourier space.
Following Refs.~\cite{book:liddle} and \cite{Malik:2006ir} we will
write $\dvp{}(k^i)$ for the Fourier component of $\dvp{}(x^i)$ such
that
\begin{equation}
 \dvp{}(\eta, x^i) = \frac{1}{(2 \pi)^3} \int \d^3k \dvp{}(\kvi) \exp (i k_i
x^i)
\,,
\end{equation}
where $\kvi$ is the comoving wavenumber.
In Fourier space, the closed form of the first order Klein-Gordon equation
\eqref{eq:KGflatsingle-num} then
transforms into
\begin{multline}
\label{eq:fokg}
 \dvp1(\kvi)'' + 2\H \dvp1(\kvi)' + k^2\dvp1(\kvi) \\
+ a^2 \left[\Upp +
\frac{\kapsq}{\H}\left(2\vp_{0}' \Uphi + (\vp_{0}')^2\frac{\kapsq
}{\H}\U\right)\right]\dvp1(\kvi) = 0 \,,
\end{multline}
where $\vp_0$ is the background homogeneous scalar field, $\dvp1$ is the first order scalar field
perturbation, $a$ is the FRW scale factor,
$\H = a'/a$ is the conformal Hubble parameter and $V$ is the potential of the scalar
field with derivatives w.r.t $\vp$ denoted by $\Vphi$ etc.
The second order equation \eqref{eq:KG2flatsingle-num} requires a careful
consideration of terms that are quadratic in the first order perturbation. 
Terms at second order of the form
$\left(\dvp1(x^i)\right)^2$ require the use
of the convolution theorem (see for example \Rref{Vretblad:2005}).
We use convolutions of the form
\begin{equation}
 f(x^i)g(x^i) \longrightarrow \frac{1}{(2 \pi)^3} \int \d^3q \d^3p\, \delta^3(\kvi
-\pvi -\qvi) f(\pvi)
g(q^i) \,.
\end{equation}
The full closed form, second order Klein-Gordon
equation in Fourier space is then given by \cite{Malik:2006ir}
\begin{align}
\label{eq:SOKG-real-num}
\dvp2''(\kvi) &+ 2\H \dvp2'(\kvi) + k^2 \dvp2(\kvi) \nonumber \\
&+ a^2\left[\Upp + \frac{\kapsq}{\H}\left(2\vp_{0}'\Uphi
+ (\vp_0')^2\frac{\kapsq}{\H}\U \right) \right]\dvp2(\kvi) \nonumber \\
&+ S(\kvi) = 0\,.
\end{align}
The source term $S$ contains all the products of $\dvp1$ in real space which require
convolution integrals. Terms which contain gradients of $\dvp1$ include additional factors of $k$
and $q$. The form of $S$ is given by \cite{Malik:2006ir}
\begin{align}
 \label{eq:Fdvk1-fourier-num}
&S\left(\dvp1(\kvi),\dvp1'(\kvi)\right)
= \frac{1}{(2\pi)^3}\int \d^3p\, \d^3q\,\delta^3(\kvi-\pvi-\qvi) 
\Bigg\{ \nonumber \\
&\quad 2\frac{\kapsq}{\H} \left[ Q \dvp1'(\pvi) \dvp1(\qvi) + \vp_{0}' a^2\Upp
\dvp1(\pvi)\dvp1(\qvi) \right]  \nonumber \\
&\quad + \left(\frac{\kapsq}{\H}\right)^2\vp_{0}'\left[2a^2\Uphi\vp_{0}'
\dvp1(\pvi)\dvp1(\qvi) + \vp_{0}'Q\dvp1(\pvi)\dvp1(\qvi) \right] \nonumber  \displaybreak[0]\\
&\quad - 2\left(\frac{\kapsq}{2\H}\right)^2\frac{\vp_{0}' Q}{\H} \left[Q\dvp1(\pvi)
\dvp1(\qvi) +
\vp_{0}' \dvp1(\pvi) \dvp1'(\qvi)\right] \nonumber \\
\biggl. &\quad + \frac{\kapsq}{2\H} \vp_{0}' \dvp1'(\pvi) \dvp1'(\qvi) 
 + a^2\left[\Uppp + \frac{\kapsq}{\H}\vp_{0}' \Upp\right] \dvp1(\pvi)
\dvp1(\qvi) \nonumber \\
&\quad + 2\left(\frac{\kapsq}{\H}\right)\frac{p_kq^k}{q^2}
\delta\vp_{1}'(\pvi)\left(Q\dvp1(\qvi)+\vp_{0}'\dvp1'(\qvi)\right)
+p^2 2\frac{\kapsq}{\H}\dvp1(\pvi)\vp_{0}'\dvp1(\qvi) \nonumber \\
&\quad 
+\left(\frac{\kapsq}{2\H}\right)^2
\frac{\vp_{0}'}{\H}\Bigg[
\left(p_lq^l-\frac{p^iq_jk^jk_i}{k^2}\right) 
\vp_{0}'\delta\vp_{1}(\pvi)\vp_{0}'\delta\vp_{1}(\qvi)
\Bigg]\nonumber \displaybreak[0]\\
&\quad +2\frac{Q}{\H}\left(\frac{\kapsq}{2\H}\right)^2 
\frac{p_lq^lp_mq^m+p^2q^2}{k^2q^2}
\Bigg[\vp_{0}'\delta\vp_{1}(\pvi)
\left(Q\dvp1(\qvi)+\vp_{0}'\dvp1'(\qvi)\right)
\Bigg]
\nonumber \displaybreak[0]\\
&\quad +\frac{\kapsq}{2\H}
\Bigg[
4Q\frac{q^2+p_lq^l}{k^2}\left(
\dvp1'(\pvi)\dvp1(\qvi)\right)
-\vp_{0}'p_lq^l \delta\vp_{1}(\pvi)\delta\vp_{1}(\qvi)
\Bigg]
\nonumber\\
&\quad +\left(\frac{\kapsq}{2\H}\right)^2
\frac{\vp_{0}'}{\H}\Bigg[
\frac{p_lq^lp_mq^m}{p^2q^2}
\left( Q\dvp1(\pvi)+\vp_{0}'\dvp1'(\pvi)\right)
\left(Q\dvp1(\qvi)+\vp_{0}'\dvp1'(\qvi)\right)
\Bigg]\nonumber \displaybreak[0]\\
&\quad +\frac{\vp_{0}'}{\H}
\Bigg[
\kapsq\left(\frac{p_lq^l+p^2}{k^2}q^2\dvp1(\pvi)\dvp1(\qvi)
-\frac{q^2+p_lq^l}{k^2}\dvp1'(\pvi)\dvp1'(\qvi)
\right)
\nonumber\\
&\quad +\left(\frac{\kapsq}{2\H}\right)^2
\frac{k^jk_i}{k^2}\Bigg(
2\frac{p^ip_j}{p^2}
\left(Q\dvp1(\pvi)+\vp_{0}'\dvp1'(\pvi)\right)
Q\dvp1(\qvi)
\Bigg)\Bigg]
\Bigg\}\,,
\end{align}
where we have defined the parameter $Q=a^2 (\kapsq \U \vp_0'/\H + \Uphi)$ for
convenience.
Calculating \eq{eq:Fdvk1-fourier-num} is the main task of the numerical simulation described in
Section~\ref{sec:code}.

% % % % % % % % % % % % % % % % % % % % % % % % % % % 
\subsection{Equations for Numerical Calculation}

In the previous section the governing equations of the second order
system were given in terms of conformal time. A more appropriate time
variable for the numerical simulation is the number of e-foldings, $\N
= \log ( a / a_{\mathrm{init}} )$.  This is measured from
$a_{\mathrm{init}}$, the value of $a$ at the beginning of the
simulation. We take the value of the scale factor today, $a_0$, as
$a_0=1$ and calculate $a$ at the end of inflation by connecting it
with $a_0$ (see discussion in \Rref{Huston:2010by} or for example Eq.~(3.19) in
\Rref{book:liddle}). We also assume that reheating was instantaneous
at the end of inflation.
We will use a dagger ($\dN{}$) to
denote differentiation with respect to $\N$.
Derivatives with respect to conformal time, $\eta$, and coordinate time, $t$, are
then given by
\begin{align}
 \pd{ }{\eta} &= \frac{\d \N}{\d \eta}\pd{}{\N} = \H \pd{}{\N} \,,\\
 \pd{ }{t} &= \frac{\d \eta}{\d t} \frac{\d \N}{\d \eta}\pd{}{\N} = H
\pd{}{\N}\,,
\end{align}
respectively, where $H = \d \ln a/\d t$ and $\H = aH$.
The background and first order equations, written in terms of the new time
variable $\N$, are
\begin{equation}
\ddN{\vp_{0}} + \left(3 + \frac{\dN{H}}{H}\right)\dN{\vp_{0}} + \frac{\Uphi}{H^2} = 0
\,,
\label{eq:bgntime}
\end{equation}
and
\begin{multline}
\ddN{\dvp1}(\kvi) + \left(3 + \frac{\dN{H}}{H}\right)\dN{\dvp1}(\kvi) 
 + \Bigg[ \left(\frac{k}{aH}\right)^2 + \frac{\Upp}{H^2} + \frac{8\pi G}{H^2}
 2\dN{\vp_{0}}\Uphi \\
+ \left.\left(\frac{8\pi G}{H}\right)^2
\left(\dN{\vp_{0}}\right)^2\U \right]\dvp1(\kvi) = 0\,. 
\label{eq:fontime}
\end{multline}
The three-dimensional convolution integral $\int \d q^3$ can be rewritten in spherical 
coordinates $q, \theta, \omega$ where $q=|\qvi|$. There is no $\omega$ dependence in the
source term integral but $|\kvi-\qvi|$ and the factors of $\kvi$ and $\qvi$ are dependent 
on $\theta$ and this dependence is made explicit below.
\eqs{eq:SOKG-real-num} and \eqref{eq:Fdvk1-fourier-num} must be written in
terms of $\N$, with the $\theta$ dependent terms grouped together, in order to set
up the numerical system completely at second order. 
In the slow-roll case there were only four different $\theta$ dependent terms,
here labelled $\A$--$\D$ following \Rref{hustonmalik}:
\begin{align}
\label{eq:AtoD-num}
 \A(\kvi,\qvi) &= \int_0^\pi \sin(\theta) \dvp1(\kvi-\qvi) \d\theta \,,
\nonumber\\
 \B(\kvi,\qvi) &= \int_0^\pi \cos(\theta)\sin(\theta) \dvp1(\kvi-\qvi)
\d\theta \,,\nonumber\\
 \C(\kvi,\qvi) &= \int_0^\pi \sin(\theta) \dN{\dvp1}(\kvi-\qvi) \d\theta \,,
\nonumber\\
 \D(\kvi,\qvi) &= \int_0^\pi \cos(\theta) \sin(\theta) \dN{\dvp1}(\kvi-\qvi)
\d\theta \,.
\end{align}
The non-slow-roll source term in \eq{eq:fullsrc-res} that we are now
considering requires the use of three further $\theta$ integrals in
addition to those in \eq{eq:AtoD-num}, which are
\begin{align}
\label{eq:efg-terms-res}
 \E(\kvi, \qvi) &= \int_0^\pi \cos^2(\theta) \sin(\theta) \dvp1(\kvi-\qvi)\d \theta
\,,\nonumber \\
\F(\kvi, \qvi) &= \int_0^\pi \frac{\sin^3(\theta)}{|\kvi-\qvi|^2} \dvp1(\kvi-\qvi)\d
\theta \,,\nonumber \\
\G(\kvi, \qvi) &= \int_0^\pi \frac{\sin^3(\theta)}{|\kvi-\qvi|^2}
\dN{\dvp1}(\kvi-\qvi)\d \theta \,.
\end{align}
It is worth noting that the term $\sin^3(\theta)/|\kvi-\qvi|^2$ tends to zero in the
limit where $k=q$ and $\theta\rightarrow 0$.
The second order Klein-Gordon equation in e-folding time is
\begin{multline}
 \label{eq:fullso-res}
\ddN{\dvp2}(\kvi) + \left(3 + \frac{\dN{H}}{H}\right) \dN{\dvp2}(\kvi) +
\left(\frac{k}{aH}\right)^2 \dvp2(\kvi) \\
+ \frac{1}{H^2}\left[ \Upp + \kapsq\left(2\dN{\vp_0}\Uphi + \kapsq
\left(\dN{\vp_0}\right)^2 \U \right)\right] \dvp2(\kvi)
+ S(\kvi) = 0\,,
\end{multline}
where the full source equation is given by
\begin{align}
\label{eq:fullsrc-res}
S(\kvi) = &\frac{1}{(2\pi)^2}\int \d q q^2 \Bigg\{ 
\frac{1}{\left(H\right)^2} \left[ \Uppp + 3(\kapsq)\dN{\vp_0}\Upp\right]
 \dvp1(\qvi) \A(\kvi, \qvi) \nonumber \\
&+\frac{(\kapsq)^2 \dN{\vp_0}}{(aH)^2}\left[ 2a^2\dN{\vp_0}\Uphi +Q \left(\dN{\vp_0}
-\frac{Q}{2(aH)^2}\right)\right] \dvp1(\qvi) \A(\kvi, \qvi) \nonumber \\
&- \frac{(\kapsq)^2}{(aH)^2}\frac{(\dN{\vp_0})^2 Q}{2} \dN{\dvp1}(\qvi)
\A(\kvi, \qvi) \nonumber \\
&+ \frac{2(\kapsq)Q}{(aH)^2} \dvp1(\qvi)\C(\kvi, \qvi) 
+ \frac{\kapsq \dN{\vp_0}}{2} \dN{\dvp1}(\qvi) \C(\kvi, \qvi) \nonumber \displaybreak[0]
\\
&+ \frac{\kapsq}{(aH)^2} \Bigg\{\dN{\vp_0}\Bigg[ \left(2k^2 + \left(\frac{7}{2} - \frac{8\pi
G}{4}(\dN{\vp_0})^2 \right)q^2 + \frac{3}{4}\frac{\kapsq}{(aH)^2} Q^2 \right)
\dvp1(\qvi) \nonumber \\
& \qquad + (\kapsq) Q \dN{\vp_0} \left(\frac{3}{4} + \frac{q^2}{k^2}\right)
 \dN{\dvp1}(\qvi) \Bigg] \A(\kvi, \qvi) \nonumber \\ \displaybreak[0]
&+ \Bigg[ \left( 2Q \frac{q}{k} \left(1- \frac{\kapsq}{(aH)^2} Q \dN{\vp_0}\right)
-\frac{9}{2} \dN{\vp_0} k q - \dN{\vp_0} \frac{q^3}{k}\right) \dvp1(\qvi) \nonumber
\\
&\qquad - 2Q (\kapsq) (\dN{\vp_0})^2\frac{q}{k} \dN{\dvp1}(\qvi)\Bigg] \B(\kvi, \qvi)
\nonumber \displaybreak[0]\\ \displaybreak[0]
&+ \Bigg[ \left(-2 + (\kapsq)(\dN{\vp_0})^2 \left(\frac{1}{4} + \frac{1}{2aH}\right)
\right) Q \dvp1(\qvi) \nonumber \\
&\qquad + \left(\frac{\kapsq}{4}(\dN{\vp_0})^2 -2\right) \dN{\vp_0} (aH)^2
\dN{\dvp1}(\qvi)\Bigg] \C(\kvi, \qvi)\nonumber \\ \displaybreak[0]
&+ \Bigg[ 2Q \frac{k}{q}\dvp1(\qvi) + \left(2\frac{k}{q}-\frac{q}{k}\right)
\dN{\vp_0} (aH)^2 \dN{\dvp1}(\qvi) \Bigg]\D(\kvi, \qvi) \nonumber \\
&+ (\kapsq) \dN{\vp_0} \Bigg[ \left(\frac{1}{4}(\dN{\vp_0})^2q^2 +
\frac{Q^2}{2(aH)^2}\right) \dvp1(\qvi) + \frac{Q}{2}\dN{\vp_0} \dN{\dvp1}(\qvi)
\Bigg] \E(\kvi, \qvi) \nonumber \displaybreak[0]\\
&+ (\kapsq)^2 \dN{\vp_0} Q \Bigg[ -\frac{Q}{2(aH)^2}\left(\frac{k^2}{2}+q^2\right)
\dvp1(\qvi) - \frac{1}{4} \dN{\vp_0} k^2 \dN{\dvp1}(\qvi) \Bigg] \F(\kvi, \qvi)
\nonumber \\
&+ (\kapsq)^2 (\dN{\vp_0})^2 \Bigg[ -\frac{Q}{2}\left(\frac{k^2}{2} +
\frac{q^2}{aH}\right)\dvp1(\qvi) -\frac{(aH)^2}{4}\dN{\vp_0}k^2 \dN{\dvp1}(\qvi)
\Bigg] \G(\kvi, \qvi) \Bigg\}\Bigg\} \,.
\end{align}
For comparison, the slow-roll expression for the source term is \cite{hustonmalik}
\begin{align}
\label{eq:KG2-source-ntime}
S(\kvi) = \frac{1}{(2\pi)^2}\int \d q\ &\Bigg\{
\frac{\Uppp}{H^2} q^2 \dvp1(\qvi) \A(\kvi,\qvi) \nonumber\\ \displaybreak[0]
&+\, \frac{8\pi G}{(aH)^2}\dN{\vp_{0}} \Bigg[ 
\left( 3a^2\Upp q^2 + \frac{7}{2}q^4 + 2k^2q^2\right) \A(\kvi,\qvi) \nonumber\\
&-\left(\frac{9}{2} + \frac{q^2}{k^2}\right)kq^3 \B(\kvi,\qvi)
\Bigg]\dvp1(\qvi) \nonumber\\ \displaybreak[0]
&+\, 8\pi G \dN{\vp_{0}} \Bigg[
-\frac{3}{2}q^2 \C(\kvi,\qvi) + \left(2-\frac{q^2}{k^2}\right)kq
\D(\kvi,\qvi) 
\Bigg]\dN{\dvp1}(\qvi) \Bigg\}\,.
\end{align}
The increased length of calculation in going from the slow-roll source term in
\eq{eq:KG2-source-ntime} to the full version in \eq{eq:fullsrc-res} can be clearly seen. The numerical
complexity is not significantly greater, however, once the three terms in \eq{eq:efg-terms-res}
have been calculated.
In the next section the implementation of the numerical scheme to calculate the source term in
\eq{eq:fullsrc-res} and solve the second order Klein-Gordon equation is discussed.

% % % % % % % % % % % % % % % % % % % % % % % % % % % % % % % % % % % % % % % % 
% 
\section{Code Implementation}
\label{sec:code}
% % % % % % % % % % % % % % % % % % % % % % % % % % % % % % % % % % % % % % % % 

% 
We described in \Rref{hustonmalik} a numerical system which calculated
the slow-roll source term as given in \eq{eq:KG2-source-ntime}. We
have improved the system and the full source term calculation has been
implemented. This code has now been released under an open source
license \cite{pyflation}.  The basic structure of the system is still
the same as outlined in \Rref{hustonmalik} and follows the form laid
down by \Rref{Salopek:1988qh,Ringeval:2007am, Martin:2006rs}. Firstly
the background fields are evolved to pinpoint the end of inflation
when $\epsilon_H=-\dN{H}/H = 1$. The first order Klein-Gordon equation
\eqref{eq:fontime} is solved using a fourth order Runge-Kutta scheme
with the initial conditions specified by the Bunch-Davies vacuum
\eq{eq:foics}.\footnote{The parameters used in the potential have previously
been fixed by fitting the first order power spectrum to the best fit
WMAP value of $\Pr=2.45\times 10^{-9}$, five e-foldings after horizon
crossing.}
Following this first order stage the source term is calculated at all
necessary timesteps.\footnote{This stage is naively parallelisable as the
calculation at each timestep is independent.  To shorten execution
time the source term can be calculated for selected $k$ modes only
instead of the full range. Note that the first order calculation still
needs to be run with a large range of $k$ modes in order that the
convolution integral can be performed.}
In the last stage of the
numerical calculation, the source term result is used to evolve the
second order perturbation modes using \eq{eq:fullso-res}.

The major advance reported in this paper is the use of the full (non-slow-roll) source equation in
the third step above. 
With this advance, seven different terms need to be calculated at each time step. These vary
in $k, q$ and $\theta$ and integrals over $q$ and $\theta$ are performed, again at each time step,
for each $k$ value. 
In order to perform these integrals we implement cutoffs at large and small values of $k$.
These cutoffs and the integration calculations in general are described in detail in
Refs.~\cite{hustonmalik,Huston:2010by} and we do not discuss them further here.
The seven different terms are functional forms of $\theta$ given in
Eqs.~\eqref{eq:AtoD-num} and \eqref{eq:efg-terms-res}. 
As the first order results are tabulated at specific $k$
values, it is necessary to interpolate the results for $k$ values
between these points.
The increased complexity of the source equation clearly increases
the number of calculations at each time step and hence the overall
execution time. The original slow-roll calculation was numerically
intensive so the effect of almost doubling the number of calculations
could have been dramatic. In fact, optimisation of the code and the
translation of some parts into compiled modules has made the increase
in complexity manageable. By selecting only particular modes of
interest, for example the WMAP pivot scale
$\kwmap=0.002\Mpc^{-1}=5.25\e{-60}\Mpl$, the source term calculation
can be significantly shortened.% 
\footnote{The run time of the source term
calculation for a single $k$ mode is approximately two hours on a
relatively modern CPU.  Using a cluster of computing nodes shortens
the naively parallelisable source term calculation to a further
degree.}

In order to solve the ODEs given in \eqs{eq:bgntime}, \eqref{eq:fontime} and \eqref{eq:fullso-res}
initial values of $\vp_0, \dvp1, \dvp2$ and their first derivatives must be specified. For the
background field initial values of $\vp_0$ and $\dN{\vp_0}$ are chosen so that the inflationary
period lasts approximately 60 e-foldings after the scales of interest in the CMB exit the horizon.
In the case of the quadratic potential this requires choosing super-Planckian initial field values,
for example $\vp_{0,\mathrm{init}}=18\Mpl$.

The first order perturbation initial conditions are chosen by assuming that sufficiently far inside
the horizon the perturbation modes are in the Bunch-Davies vacuum state. To implement this early
time condition in the numerical system we follow Salopek \etal \cite{Salopek:1988qh} who initialise
the first order modes at a time when $k/aH$ for the mode equals some arbitrary factor. In keeping
with \Rref{Salopek:1988qh} we choose this factor to be $50$. The first order initial values are
then calculated as
\begin{align}
\label{eq:foics}
 \dvp1|_{\mathrm{init}} &= \frac{\sqrt{8\pi G}}{a}
\frac{e^{-i k\eta}}{\sqrt{2k}} \,,\\
 \dN{\dvp1}|_{\mathrm{init}} &= -\frac{\sqrt{8\pi G}}{a}
\frac{e^{-i k\eta}}{\sqrt{2k}} \left(1 + i \frac{k}{a H}\right) \,,
\end{align}
where $\eta$ is the conformal time again.

The situation of the second order initial conditions is different. At
the initial times when the Bunch-Davies conditions are suitable the
perturbations are expected to be highly Gaussian.  In this paper we
are interested in the production of second order effects by the
evolution of the first order modes and we make no assumptions about
the existence of second order perturbations before the simulation
begins. Therefore, we set the initial condition for each second order
perturbation mode to be $\dvp2=0$, and $\dN{\dvp2}=0$ at the time when
the corresponding first order perturbation is initialised.

A numerical solution for the second order perturbation equation will contain a homogeneous solution
and a particular solution. As stated
above we have chosen the initial values for the second order field to be
zero. On their own these initial conditions do not remove this homogeneous solution from the result
for $\dvp2$ in general. In order to do this, and keep only the 
particular solution to the equation, it is necessary to ensure that the homogeneous solution is the
trivial $(0,0)$ solution throughout the evolution.

\begin{figure}
 \centering
\subfloat[The ramp function used to remove the homogeneous solution for $\dvp2$, here
shown around the initialisation time for the scale \kwmap.]{
 \includegraphics[width=\twofigwidth]{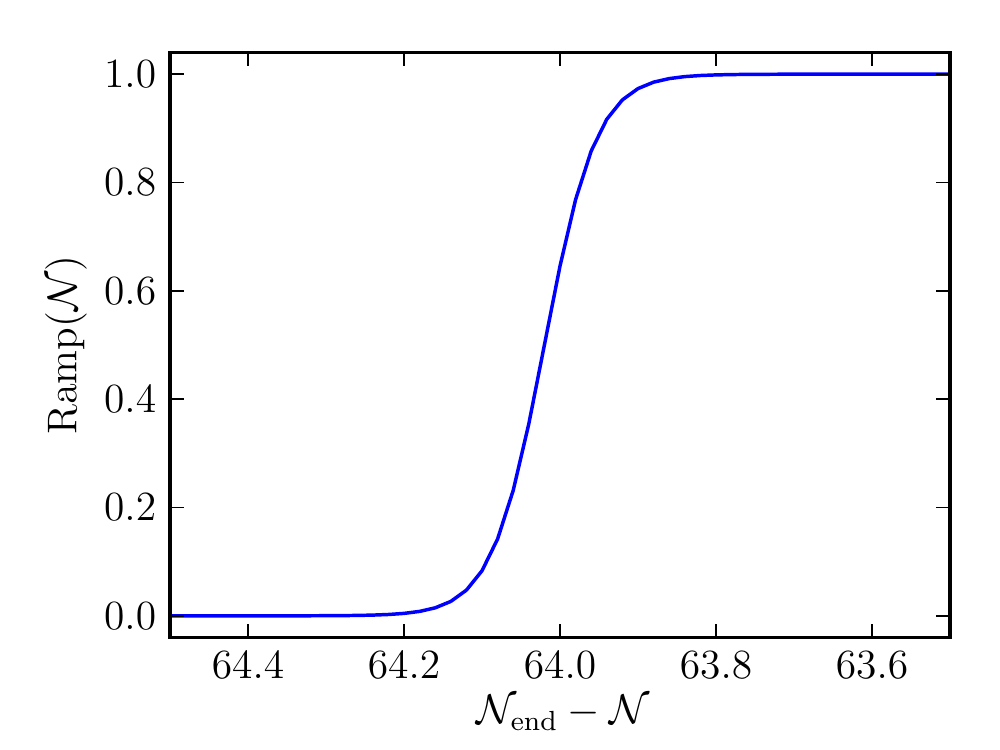}
 % ramp-kwmap-large.pdf: 432x324 pixel, 72dpi, 15.24x11.43 cm, bb=0 0 432 324
 \label{fig:ramp}
}
% \hspace{5mm}
% 
\subfloat[The addition of the ramp term to $S$ changes its value at the initialisation time from
the original value (blue solid line) to the ramped value (green dashed line), here
shown for scale $\kwmap$.]{
 \includegraphics[width=\twofigwidth]{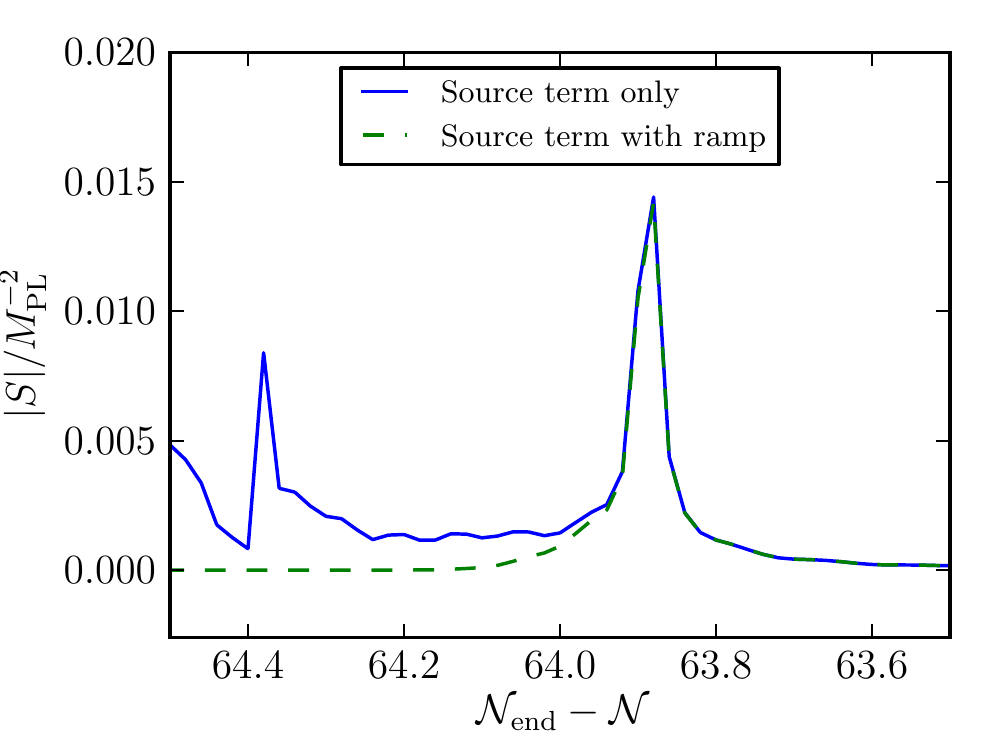}
 \label{fig:src-with-ramp}
}
\caption{The ramping function used with the source term results.}
\label{fig:ramp-both}
\end{figure}

In order to only report the particular solution of the second order differential equation
\eq{eq:fullso-res} we have added a ramping term to the source term $S$. This ramp interpolates
between $0$ and $1$ for about an e-folding of time around the time of the initialisation of the
modes.
By starting the solution of $\dvp2, \dN{\dvp2}$ at $0,0$ and setting the source term to $0$
at this time through the use of the ramp, the solution for $\dvp2$ will consist only of the
inhomogeneous part. The homogeneous solution of \eq{eq:fullso-res} \iec the solution without the
source term present, is the same form as the solution for $\dvp1$ as can been seen by comparing
Eqs.~\eqref{eq:fullso-res} and \eqref{eq:fontime}. Figure~\ref{fig:ramp-both} shows the effect
that incorporating the ramp has on the source term at early times for the $\kwmap$ mode. The value
of $\dvp2(\kwmap)$ is initialised to be equal to zero at 64.3 e-foldings before the end of
inflation, when the ramp value is still zero. The ramp is added to the source term for each mode at
the respective initialisation time and all the results below were generated using a ramped
source term. 

% % % % % % % % % % % % % % % % % % % % % % % 
\section{Results}
\label{sec:results}

\subsection{Comparison with Slow-Roll Results}
\begin{figure}
\centering
\subfloat[The source term results at all time steps]{
 \includegraphics[width=\twofigwidth]{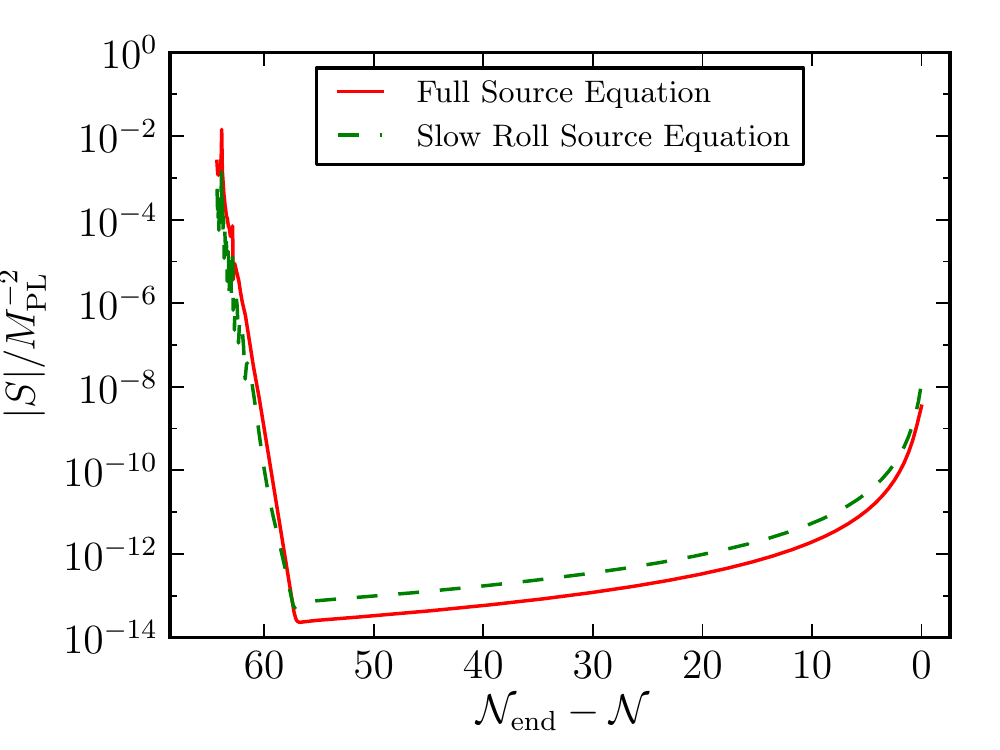}
\label{fig:src-fullandslowroll-comparison}
}
\centering
\subfloat[The second order results around the feature]{
 \includegraphics[width=\twofigwidth]{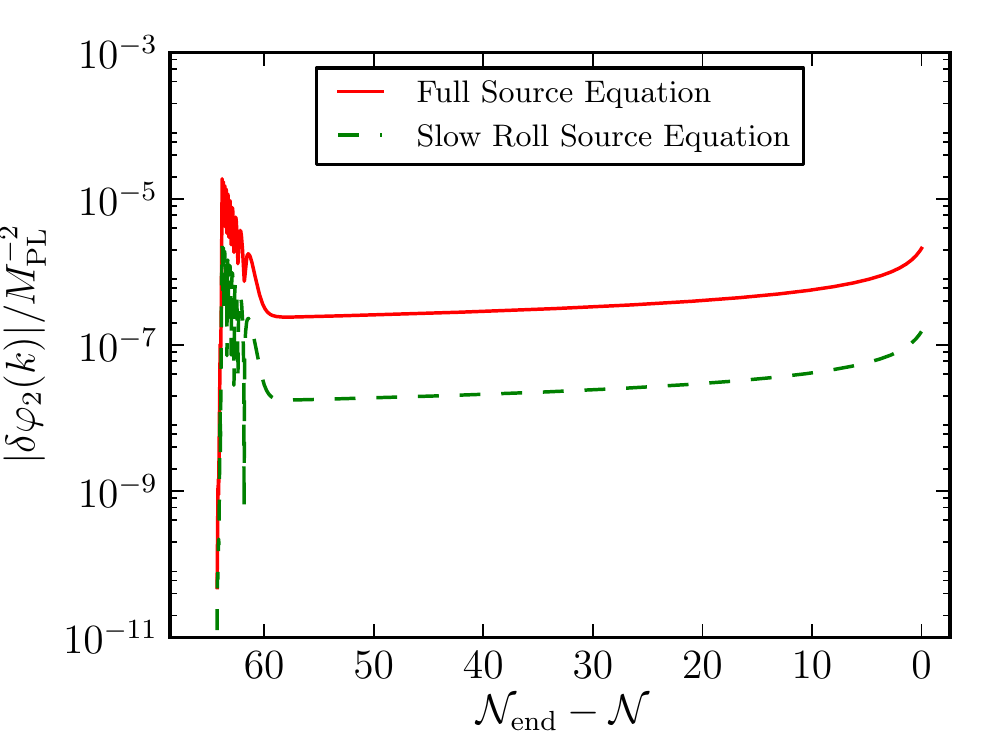}
\label{fig:dp2-fullandslowroll-comparison}
}
\caption{A comparison of the results from the full equations (red solid line) and the slow-roll
source term (green dashed line) for scale $\kwmap$.}
\label{fig:fullandslowroll-both}
\end{figure}

The full second order code has been run with a number of different potentials. Firstly in order to
check the consistency of the full equation, the standard slow-roll quadratic potential was used. 
Figure~\ref{fig:src-fullandslowroll-comparison} shows that the results of the full system are very
similar to the slow-roll results, as
expected for this potential. The additional terms in the source equation \eq{eq:fullsrc-res} subdue
some of the oscillatory noise evident in the slow-roll solution at early times when the mode is
inside the horizon. 

% % % % % % % % % % % % 
% Figure for first order power spectrum of full, half and zero step.
% % % % % % % % % % % % % 

The second order result is similar in both cases, however there is an appreciable increase in the
amplitude of the second order modes when the full equations are used. This is likely to be a
result of the reduced
oscillations mentioned above. The second order values are plotted in
Figure~\ref{fig:dp2-fullandslowroll-comparison}.

% % % % % % % % % % % % 
% Figure for source term of perturbed msqphisq potentials
\begin{figure}
\centering
\subfloat[The source term results for the perturbed models]{
 \includegraphics[width=\twofigwidth]{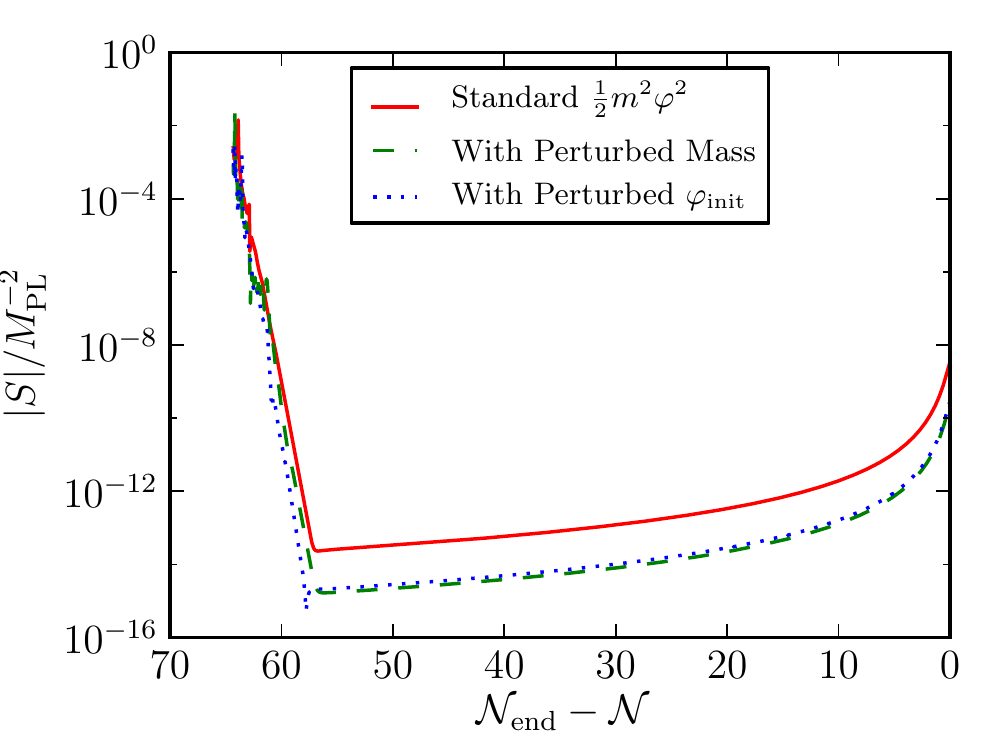}
\label{fig:src-perturb-comparison}
}
\centering
\subfloat[The second order perturbations for the perturbed models]{
 \includegraphics[width=\twofigwidth]{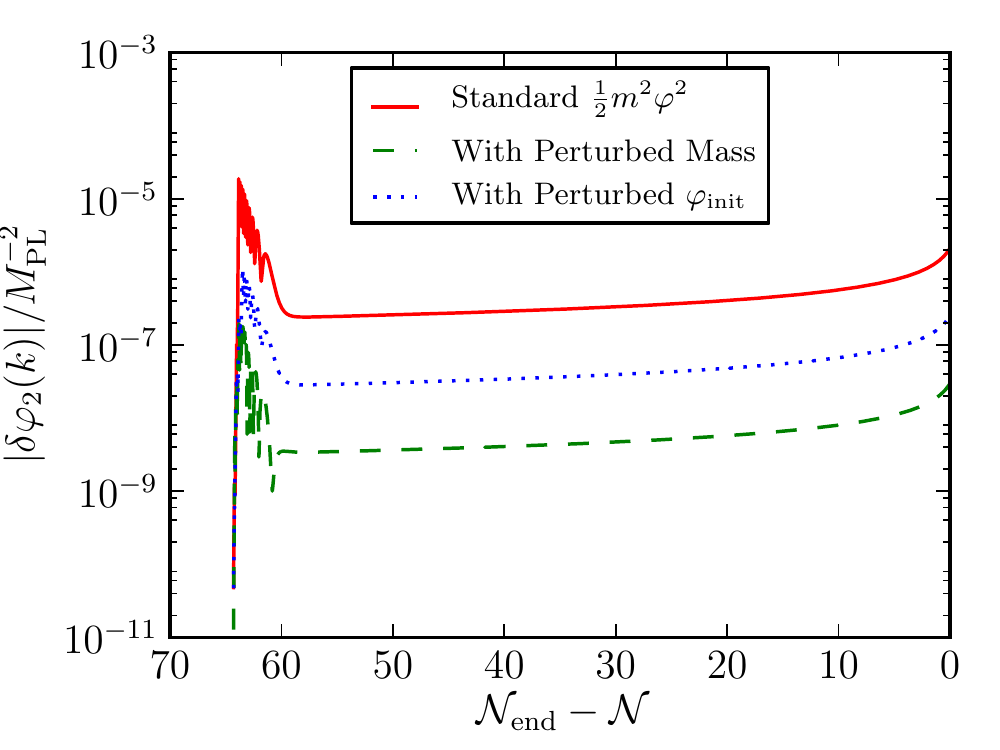}
\label{fig:src-perturb-comparison-zoom}
}
\caption{The source term and second order results for models with slightly perturbed mass (green
dashed line) and $\vp_\mathrm{init}$ (blue dotted line) compared to the standard quadratic model
(red solid line) for scale $\kwmap$.}
\label{fig:src-perturb-both}
\end{figure}
% % % % % % % % % % % % % 

The results for background and first order perturbations are robust under small deviations in the
initial conditions or the mass parameter in the potential. However, the source term and second
order results are not so robust, with any slight variation in the initial value of $\vp_0$ or the
mass parameter translating to sizable changes in the magnitude of the results.
For example, a small change in the initial field value of $\Delta\vp_{0,\mathrm{init}}=0.01\Mpl$ or
in the mass of the inflaton of $\Delta m = 1\e{-11}\Mpl$ leads to large differences in the source
term and second order results and Figure~\ref{fig:src-perturb-both} shows the effects of these
changes.

\subsection{Step and Bump Potentials}
Beyond the standard quadratic model, a more interesting potential to
consider is one with a feature at a particular scalar field
value \cite{Adams:2001vc}. Following Chen \etal \cite{Chen:2008wn, Chen:2006xjb} we have used both
a step and a bump potential. The step potential is a modified
$\msqphisq$ potential of the form
\begin{equation}
\label{V_bump}
 V_{\rm{s}}(\vp) = \msqphisq \left[
1 + c\left(\tanh\left(\frac{\vp-\vp_{\rm{s}}}{d}\right) 
- 1 \right) \right]\,,
\end{equation}
and the bump potential is given by
\begin{equation}
 V_{\rm{b}}(\vp) = \msqphisq  \left[1 + c\sech\left(\frac{\vp-\vp_{\rm{b}}}{d}\right) \right]\,,
\end{equation}
where the parameters $c,d, \vp_{\rm{s}}$ and $\vp_{\rm{b}}$ control the height,
width and central point of the feature respectively. The step
potential used here has an extra $(-1)$ term compared to the one used
in \Rref{Chen:2008wn}. This extra term ensures that
$V_{\rm{s}}\rightarrow\msqphisq$ at early times instead of beginning with a
greater amplitude than the standard quadratic potential and only
equalling the quadratic value exactly at the feature point. The
evolution of the Hubble parameter $H$ is different for both the bump
and step potentials compared to the quadratic potential. This changes
the value of $a$ at the end of inflation and therefore the value
$a_\mathrm{init}$ used at the start of the run. In order to compare
models with similar initial conditions the following results are for
runs where $a_\mathrm{init}$ has been made equal in each case. In
physical terms this amounts to small redefinitions of the value of $a$
today away from unity, or the consideration of slightly different
physical scales today. Due to the strong dependence on initial
conditions as shown above, comparison of numerical results is better
facilitated by this fixing of $a_\mathrm{init}$ than comparing models
with different values of $a_\mathrm{init}$.

% % % % % % % % % % % % 
% Figure for potential of step and bump
\begin{figure}
\centering
\subfloat[The potential $V(\vp)$ for the step, bump and
standard quadratic potentials around the feature.]{
\includegraphics[width=\twofigwidth]{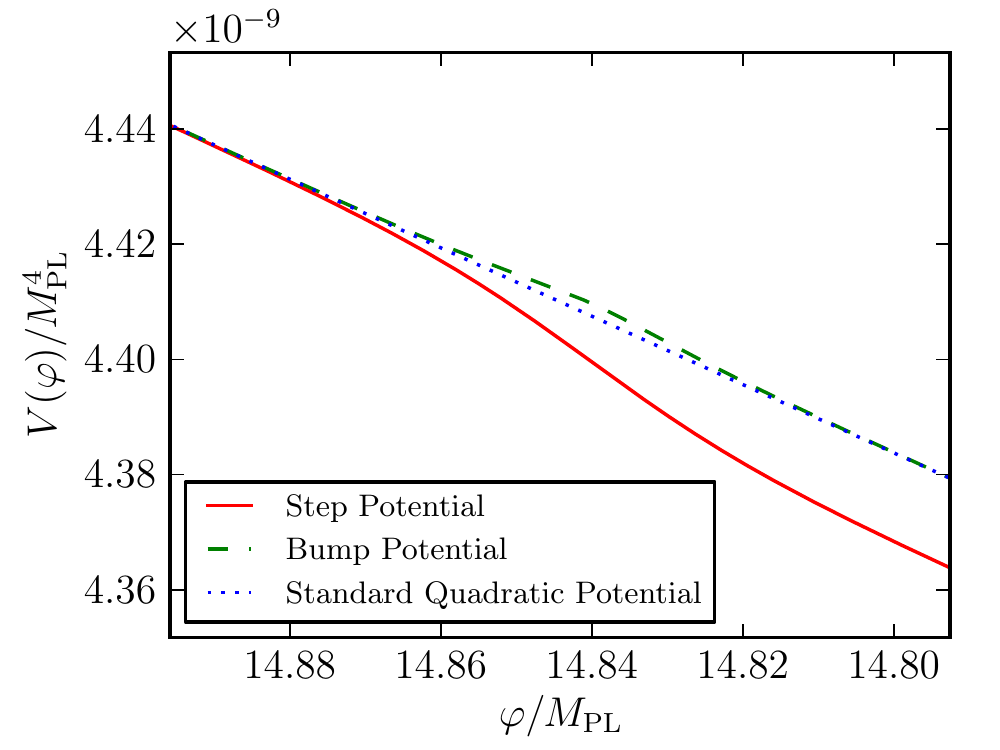}
\label{fig:pot-stepbump-zoom}
}
\subfloat[The slow-roll parameter $\eta_V$ for the step, bump and quadratic
potentials.]{
 \includegraphics[width=\twofigwidth]{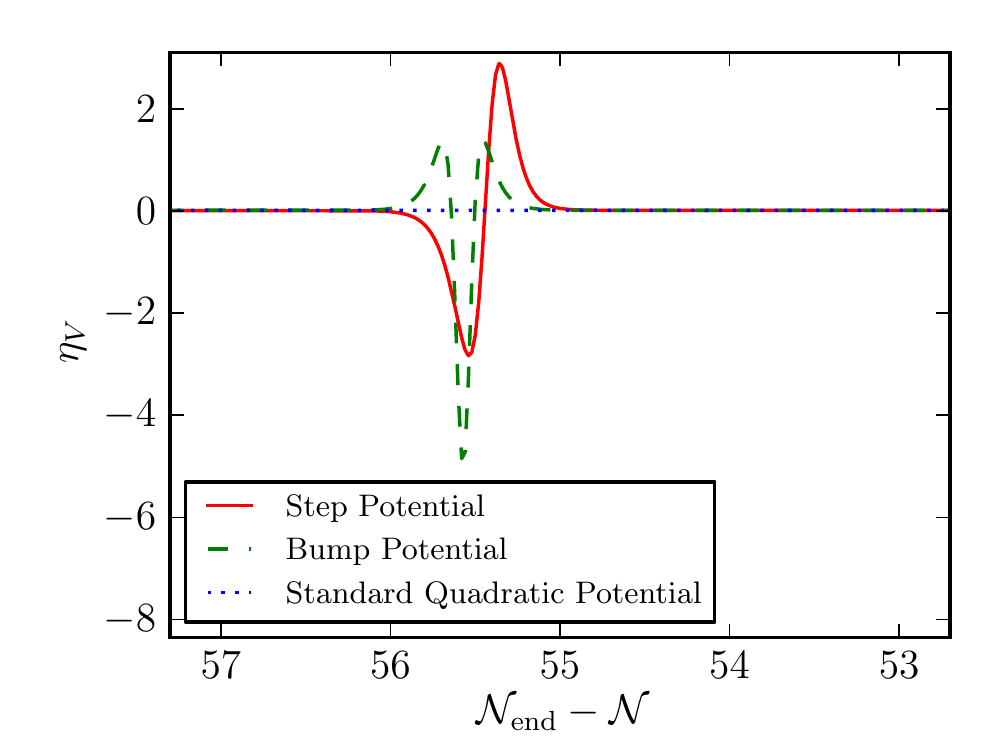}
\label{fig:eta-stepbump-zoom}
}
\caption{The potential and $\eta_V$ slow-roll parameter for the step (red solid line), bump (green
dashed line) and quadratic (blue dotted line) potentials.}
\label{fig:stepbump-potandeta}
\end{figure}
% % % % % % % % % % % % % 

% % % % % % % % % % % % %
% Table of parameters in step and bump.
\begin{table}
\centering
\begin{tabular}{ccc}
\toprule
 & Step & Bump \\
\midrule 
$c$ & 0.0018 & 0.0005 \\ 
$d$ & 0.022 & 0.01 \\ 
$\vp_{\rm{s,b}}$ & 14.84 \Mpl & 14.84 \Mpl\\
\bottomrule
\end{tabular}
\caption{The values of the parameters in the step and bump potentials.}
\label{tab:step-bump-params}
\end{table}
Figure~\ref{fig:pot-stepbump-zoom} shows the step and bump potentials at the relevant $\vp$ values.
The features are quite shallow
for this choice of parameters but can be tuned to be stronger. The values used in the code are
given in Table~\ref{tab:step-bump-params}. 
With these values the slow-roll approximation is temporarily violated as $|\eta|$ gets large around
the feature as shown in Figure~\ref{fig:eta-stepbump-zoom}.
We have run the step potential model with a ``full'' step, corresponding to $c=0.0018$, a ``half''
step with $c=0.0009$ and finally with $c$ set equal to zero.
As a ``sanity check'' the $c=0$ run gives back the same results as the standard quadratic potential
with no feature. The bump potential model has also been run with a ``full'' bump for which
$c=0.0005$, a ``half'' bump with $c=0.00025$ and a ``zero'' bump.

% % % % % % % % % % % % 
% Figure for first order power spectrum of full, half and zero step.
\begin{figure}
\centering
\subfloat[The first order results at all time steps]{
 \includegraphics[width=\twofigwidth]{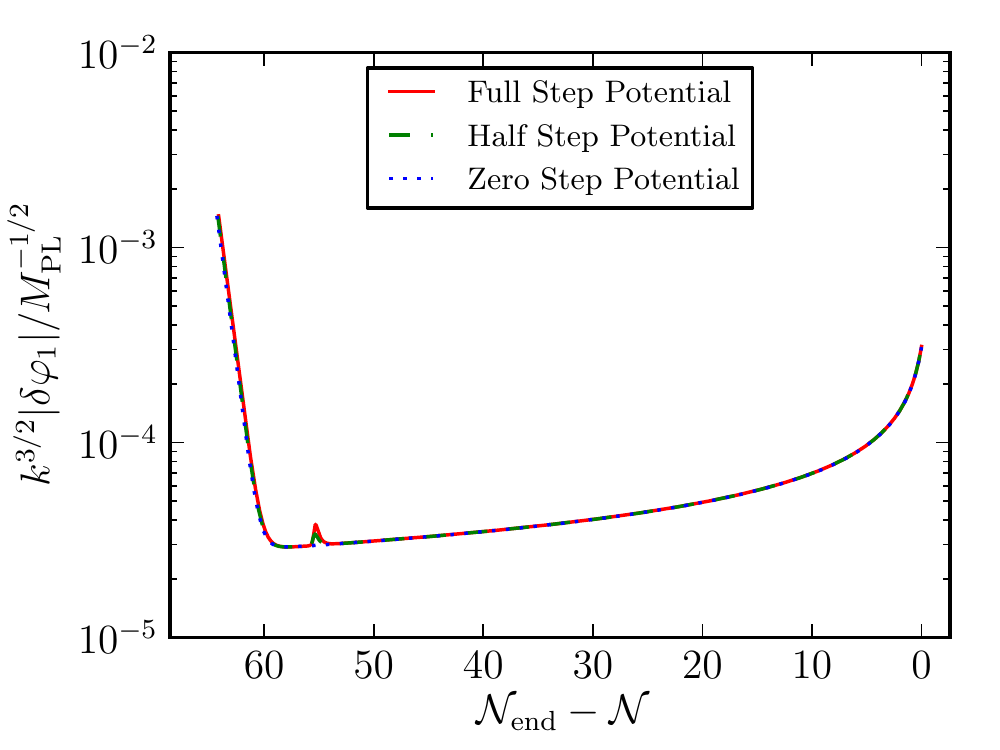}
\label{fig:dp1-step-comparison}
}
\centering
\subfloat[The first order results around the feature]{
 \includegraphics[width=\twofigwidth]{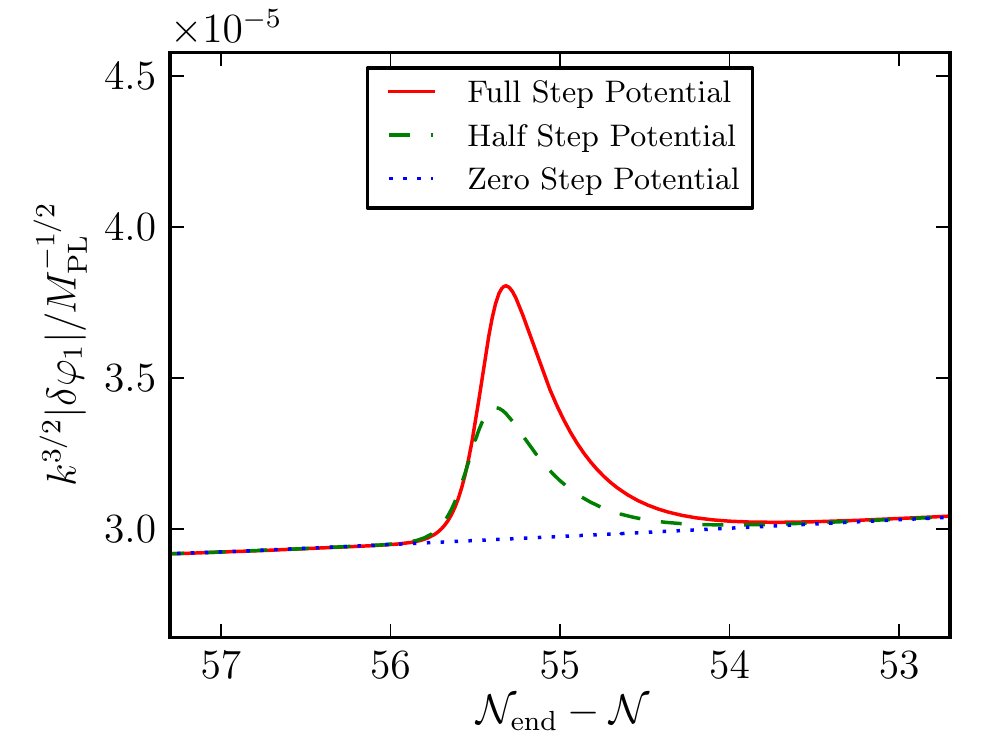}
\label{fig:dp1-step-comparison-zoom}
}
\caption{The first order results for the full (red solid line), half (green dashed line) and zero
(blue dotted line) step potentials for scale $\kwmap$.}
\label{fig:dp1-step-both}
\end{figure}
% % % % % % % % % % % % % 
% % % % % % % % % % % % 
% Figure for first order power spectrum of full, half and zero bump.
\begin{figure}
\centering
\subfloat[The first order results at all time steps]{
 \includegraphics[width=\twofigwidth]{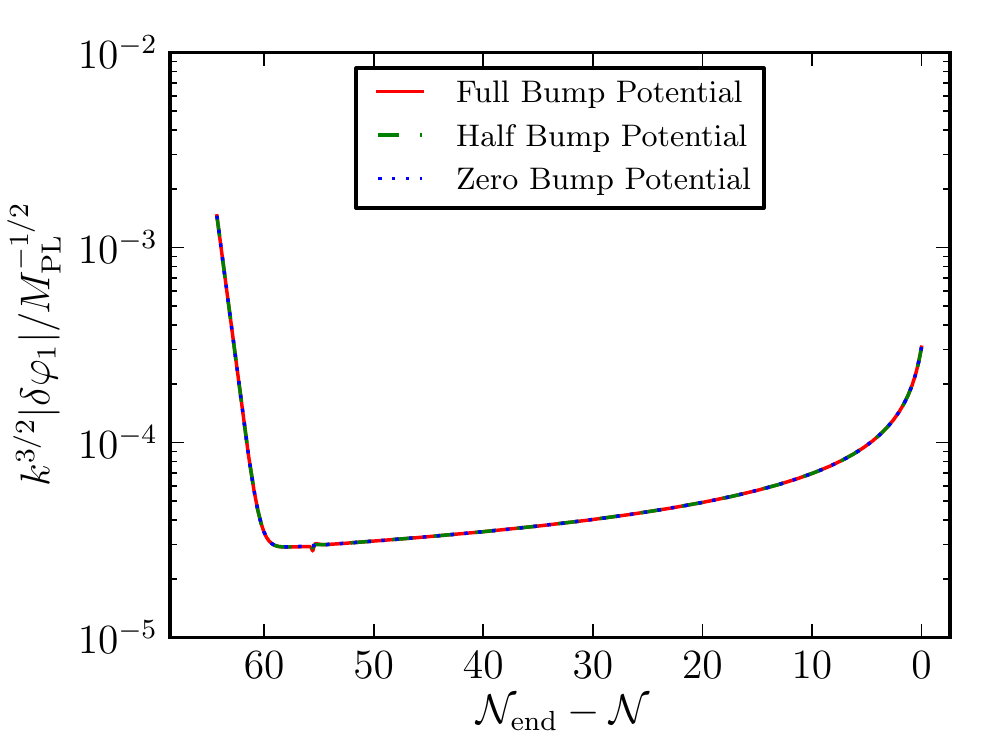}
\label{fig:dp1-bump-comparison}
}
\centering
\subfloat[The first order results around the feature]{
 \includegraphics[width=\twofigwidth]{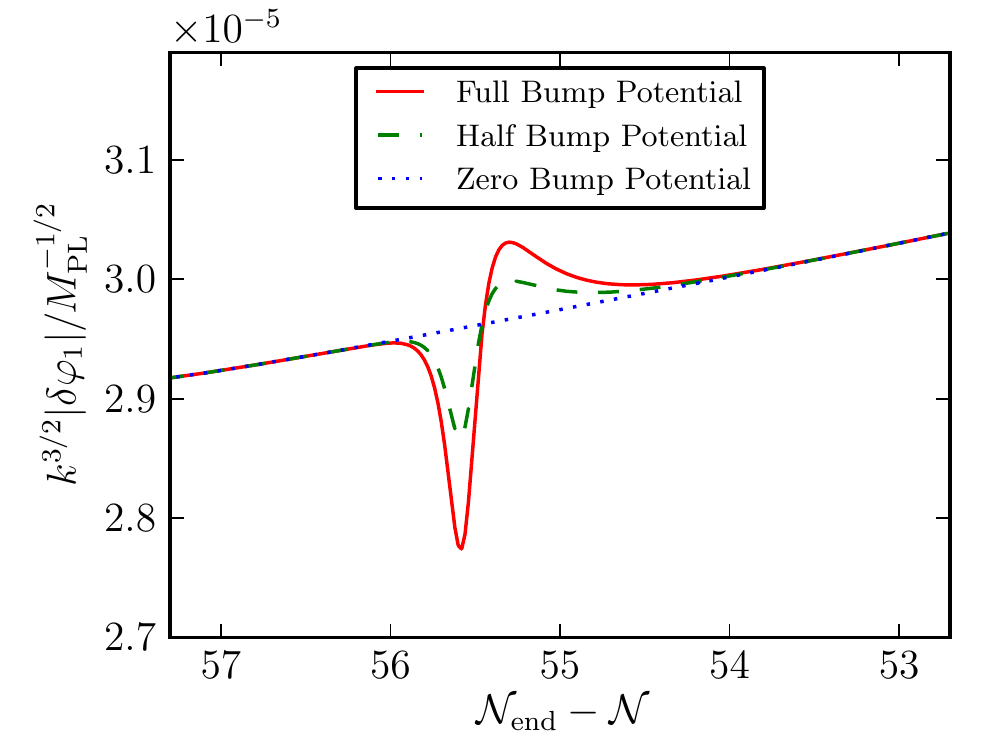}
\label{fig:dp1-bump-comparison-zoom}
}
\caption{The first order results for the full (red solid line), half (green dashed line) and zero
(blue dotted line) bump potentials for scale $\kwmap$.}
\label{fig:dp1-bump-both}
\end{figure}
% % % % % % % % % % % % % 

\begin{figure}
\centering
\subfloat[The source term results at all time steps]{
 \includegraphics[width=\twofigwidth]{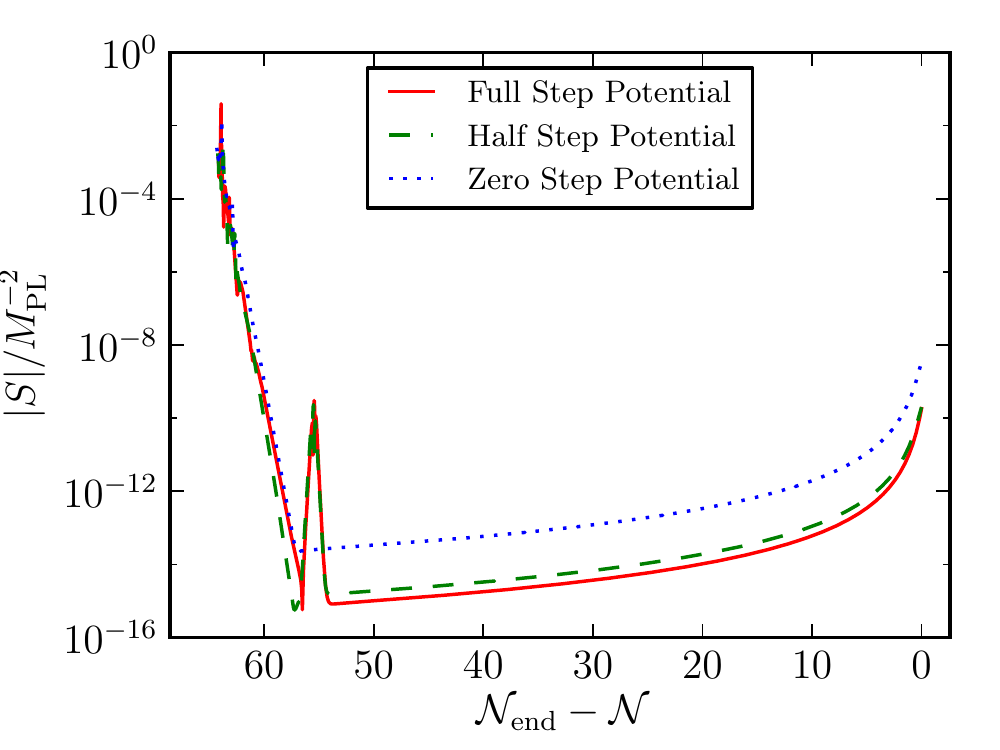}
\label{fig:src-step-comparison}
}
\centering
\subfloat[The source term results around the feature]{
 \includegraphics[width=\twofigwidth]{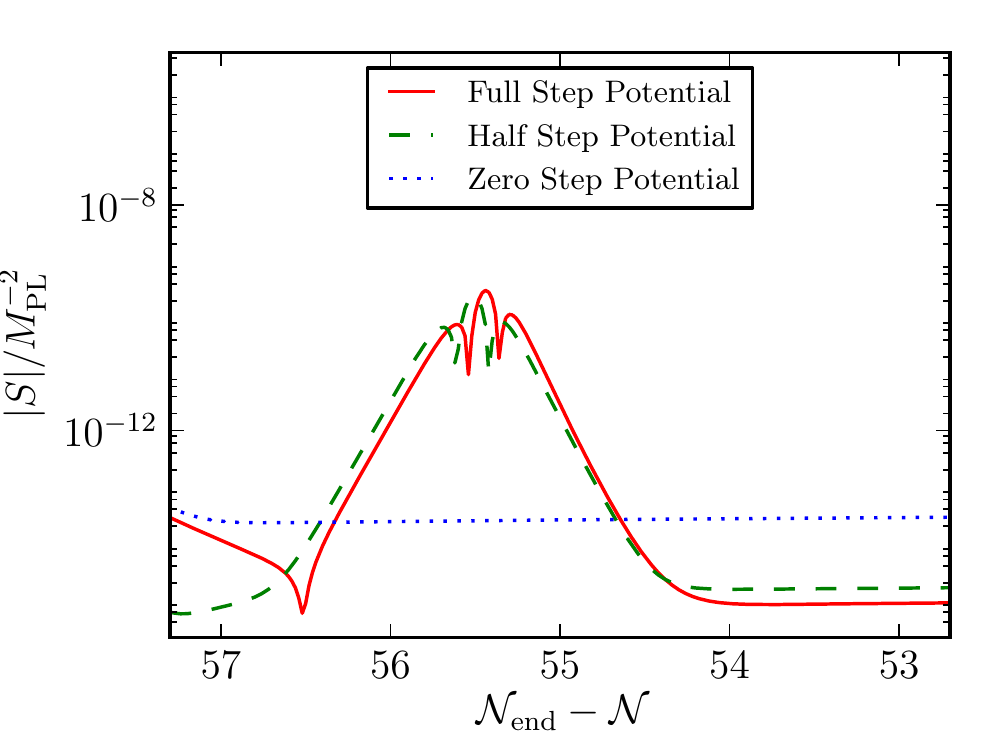}
\label{fig:src-step-comparison-zoom}
}
\caption{The source term results for the full (red solid line), half (green dashed line) and zero
(blue dotted line) step potentials for scale $\kwmap$.}
\label{fig:src-step-both}
\end{figure}

\begin{figure}
\centering
\subfloat[The source term results at all time steps]{
 \includegraphics[width=\twofigwidth]{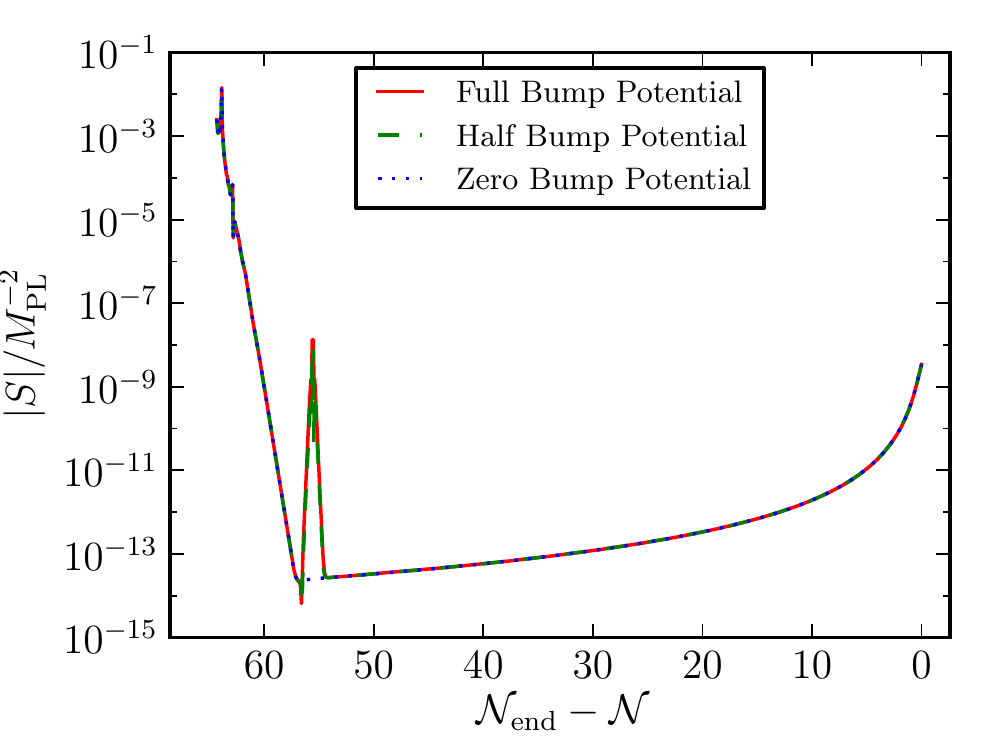}
\label{fig:src-bump-comparison}
}
\centering
\subfloat[The source term results around the feature]{
 \includegraphics[width=\twofigwidth]{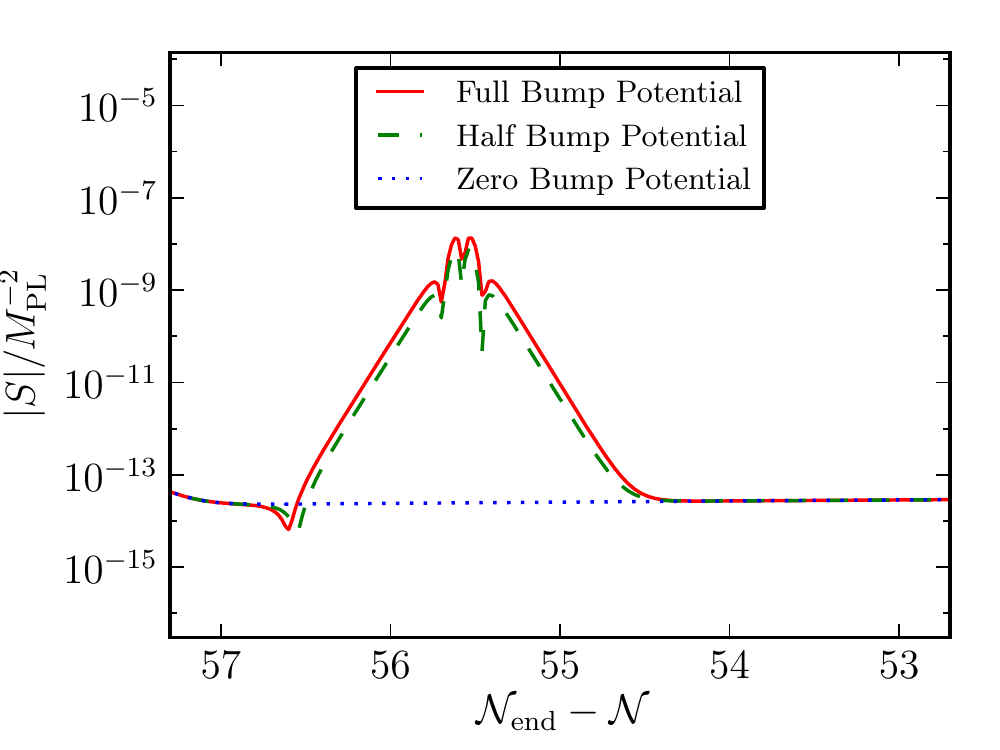}
\label{fig:src-bump-comparison-zoom}
}
\caption{The source term results for the full (red solid line), half (green dashed line) and zero
(blue dotted line) bump potentials for scale $\kwmap$.}
\label{fig:src-bump-both}
\end{figure}

The first order spectrum for the three cases for the step potential is plotted in
Figure~\ref{fig:dp1-step-both} and the cases for the bump potential in
Figure~\ref{fig:dp1-bump-both}. At first
order the effect of the feature in both potentials is localised around a particular $\vp$ value and
therefore $\N$ value. The half step and bump deviations are smaller than the full ones,
however the ratio of the amplitude changes are not symmetric around the central point of the
feature. 
In the source term calculation the presence of a feature makes a great difference in the result.
Figures~\ref{fig:src-step-both} and \ref{fig:src-bump-both} compare the results for the step and
bump potentials again with a full, half
and zero feature. For the step potential the magnitude of the source term deviates from the
standard quadratic result quite far in advance of the feature. Interestingly the change in
magnitude around the feature is almost equal in the full and half cases and is certainly not
proportional to the parameter $c$ in the way the first order results are. The magnitude of the
source term is reduced compared to the quadratic result and this reduction continues beyond the
feature. This reduction occurs before the feature when the step potential should be well
approximated by the quadratic one. The higher order derivatives will be different however and in
particular $\Vppp$ for the step potential is given by
\begin{align}
\Vppp = \frac{c m^2}{2 d^3} \sech^4\left[\frac{\vp_0-\vp_{\rm{s}}}{d}\right] &\left(3 d^2-4
\vp_0^2+\left(3 d^2+2 \vp_0^2\right) \cosh\left[\frac{2 (\vp_0-\vp_{\rm{s}})}{d}\right] \right.\\ \nonumber
& \left. -\,6 d \vp_0 \sinh\left[\frac{2 (\vp_0-\vp_{\rm{s}})}{d}\right] \right) \,,
\end{align}
where $\vp_{\rm{s}}$ is the value of the field at the feature. Unlike the third derivative of the
quadratic potential this is not zero everywhere. 
The effect on the source term can be understood by examining \eq{eq:fullsrc-res} and observing that
the first term is proportional to $\Vppp$. In addition, the second term in the equation is
proportional to $\Vpp$ and will not be constant as in the quadratic case.
In contrast, the results for the bump potential are not affected beyond the small region around the
feature, as shown in Figure~\ref{fig:src-bump-both}. Again the change in magnitude does not seem to
be proportional to the parameter $c$. Before and after the feature the result is indistinguishable
from the quadratic case even though higher order derivatives of the potential are different to the
values for the quadratic potential.

% % % % % % % % % % % % 
% Figure for source term of full, half and zero step potentials.
% % % % % % % % % % % % % 
% % % % % % % % % % % % 
% Figure for source term of full, half and zero bump potentials.
% % % % % % % % % % % % % 

\begin{figure}
\centering
\subfloat[The second order results at all time steps]{
 \includegraphics[width=\twofigwidth]{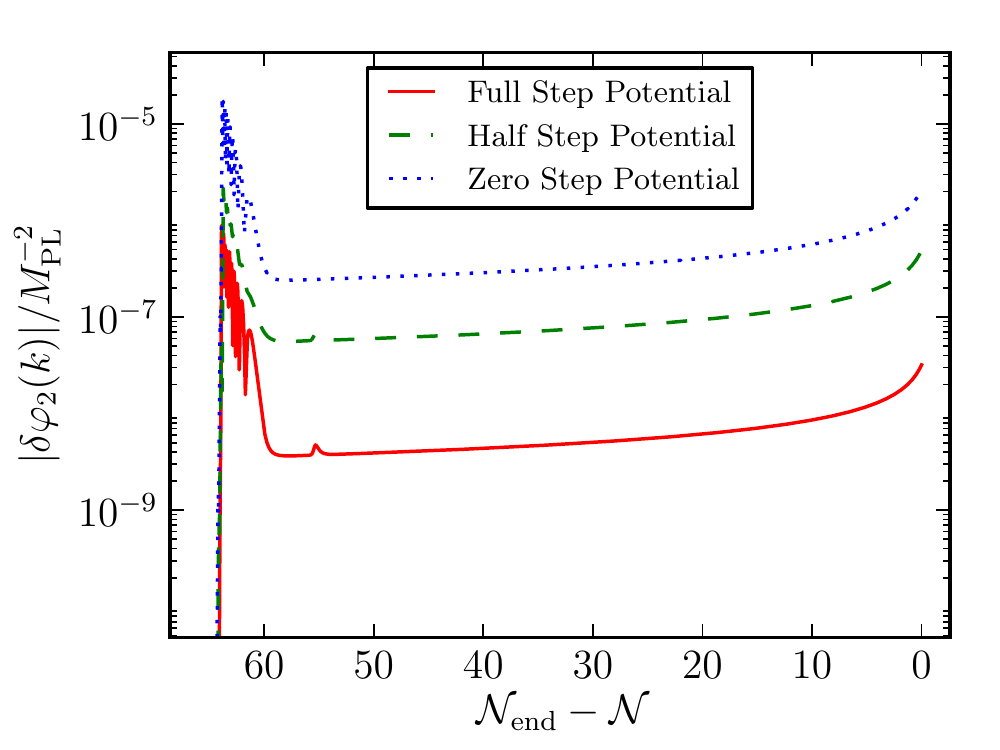}
\label{fig:dp2-step-comparison}
}
\centering
\subfloat[The second order results around the feature]{
 \includegraphics[width=\twofigwidth]{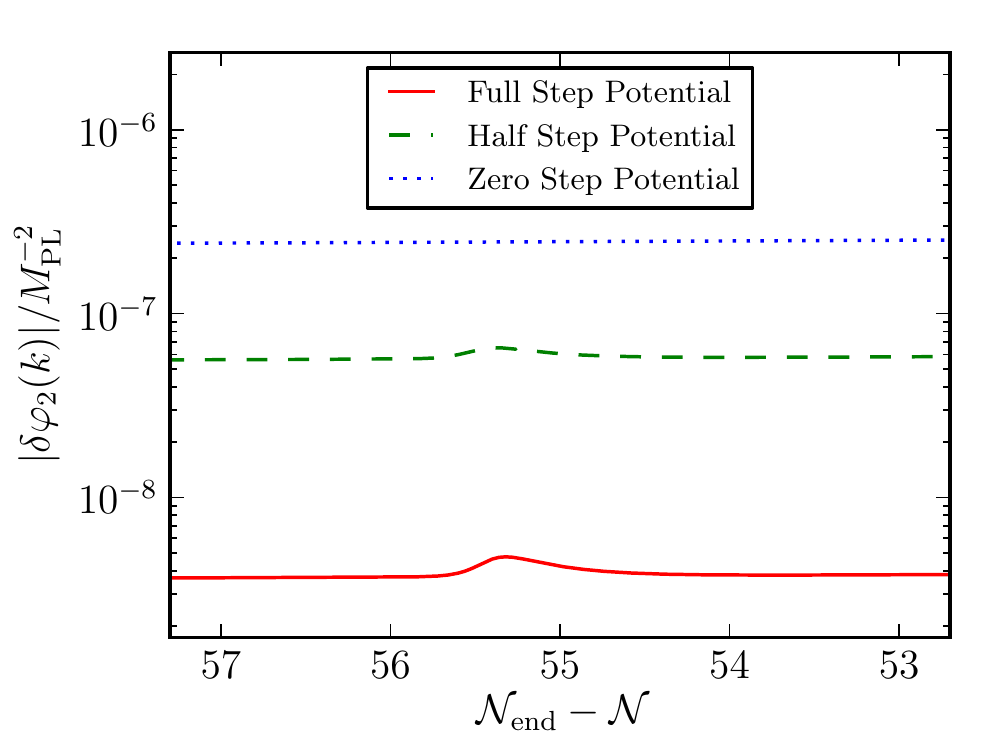}
\label{fig:dp2-step-comparison-zoom}
}
\caption{The second order results for the full (red solid line), half (green dashed line) and zero
(blue dotted line) step potentials for scale $\kwmap$.}
\label{fig:dp2-step-both}
\end{figure}

\begin{figure}
\centering
\subfloat[The second order results at all time steps]{
 \includegraphics[width=\twofigwidth]{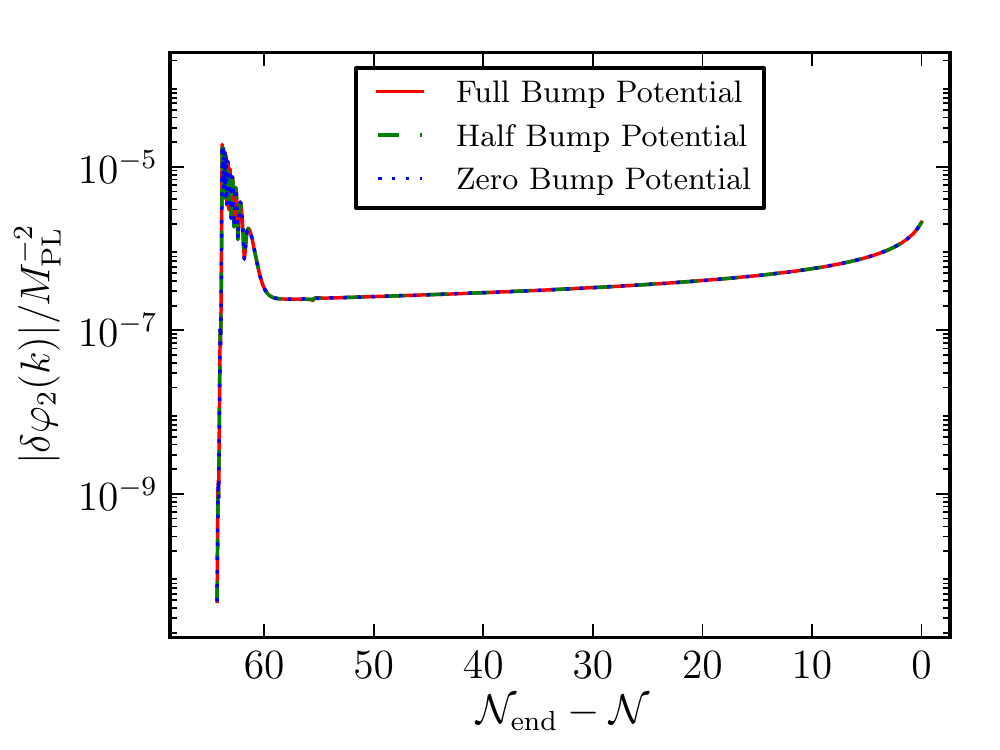}
\label{fig:dp2-bump-comparison}
}
\centering
\subfloat[The second order results around the feature]{
 \includegraphics[width=\twofigwidth]{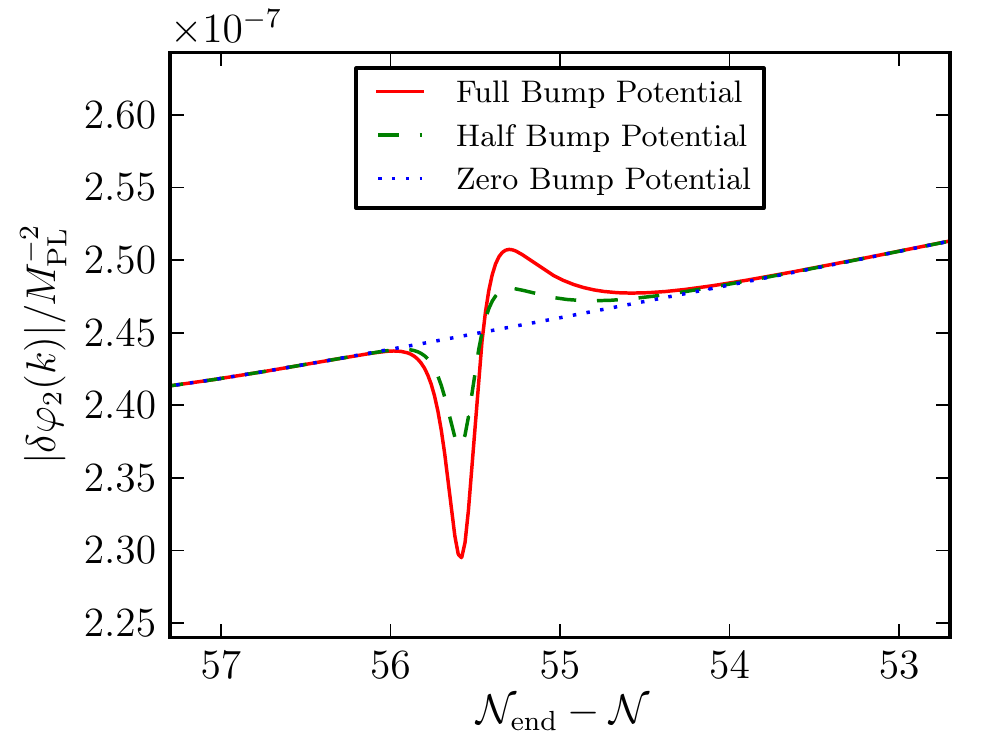}
\label{fig:dp2-bump-comparison-zoom}
}
\caption{The second order results for the full (red solid line), half (green dashed line) and zero
(blue dotted line) bump potentials for scale $\kwmap$.}
\label{fig:dp2-bump-both}
\end{figure}

% 
At second order the differences in the models make themselves even
more apparent. Figure~\ref{fig:dp2-step-both} shows the second order
solution for $\kwmap$ again for the full, half and zero step
potentials.  The amplitude of the step has a marked effect on the
amplitude of the second order modes. The differences in the source
term before the feature are carried over to the second order
result. The magnitude of the second order result is much lower for the
full step potential than the the quadratic result. When the amplitude
of the step is halved the change in the magnitude is also reduced. The
difference between the full and quadratic results is at least an order
of magnitude and the detailed cause of this difference is a subject
for further investigation.
The results for the bump potential are shown in Figure~\ref{fig:dp2-bump-both}. Here the effect of
the feature is only evident around the bump, again carrying over the result from the source term 
values. 
% % % % % % % % % % % % 
% Figure for second order power spectrum of full, half and zero step.
% % % % % % % % % % % % %
% % % % % % % % % % % % 
% Figure for second order power spectrum of full, half and zero bump.
% % % % % % % % % % % % %

\subsection{Sub- and Super-Horizon Features}
% % % % % % % % % % % % 
% Figure for first order perturbation of the bump potential for sub and super horizon features.
\begin{figure}
\centering
\subfloat[The first order results around the feature.]{
 \includegraphics[width=\twofigwidth]{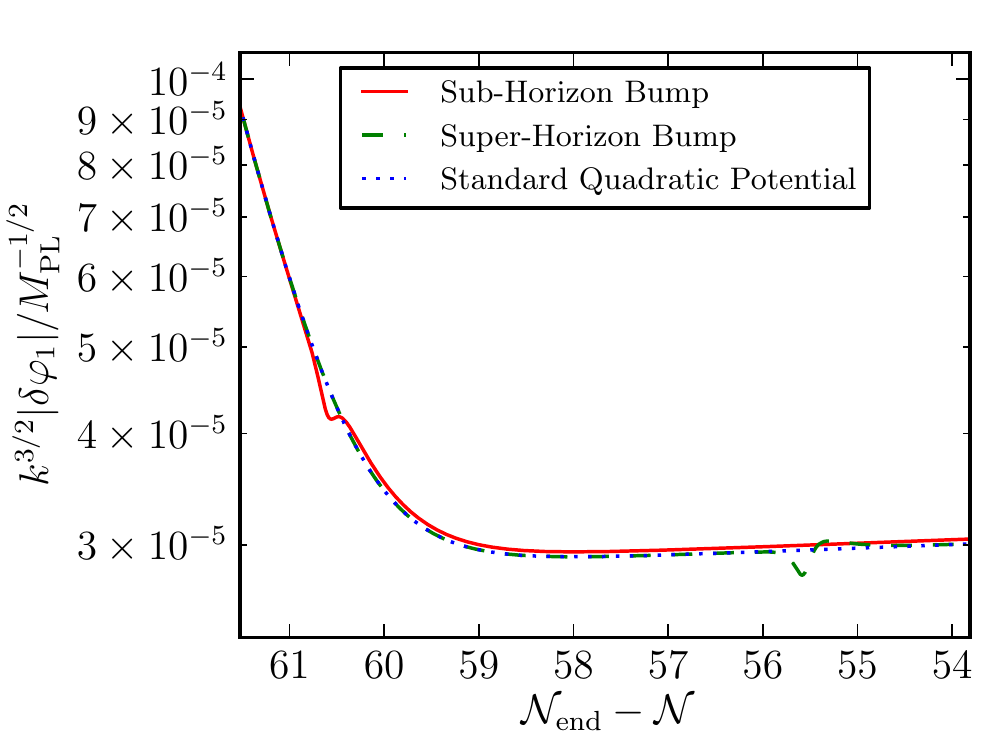}
\label{fig:dp1-bump-subandsuper}
}
\subfloat[The source term results around the feature.]{
\includegraphics[width=\twofigwidth]{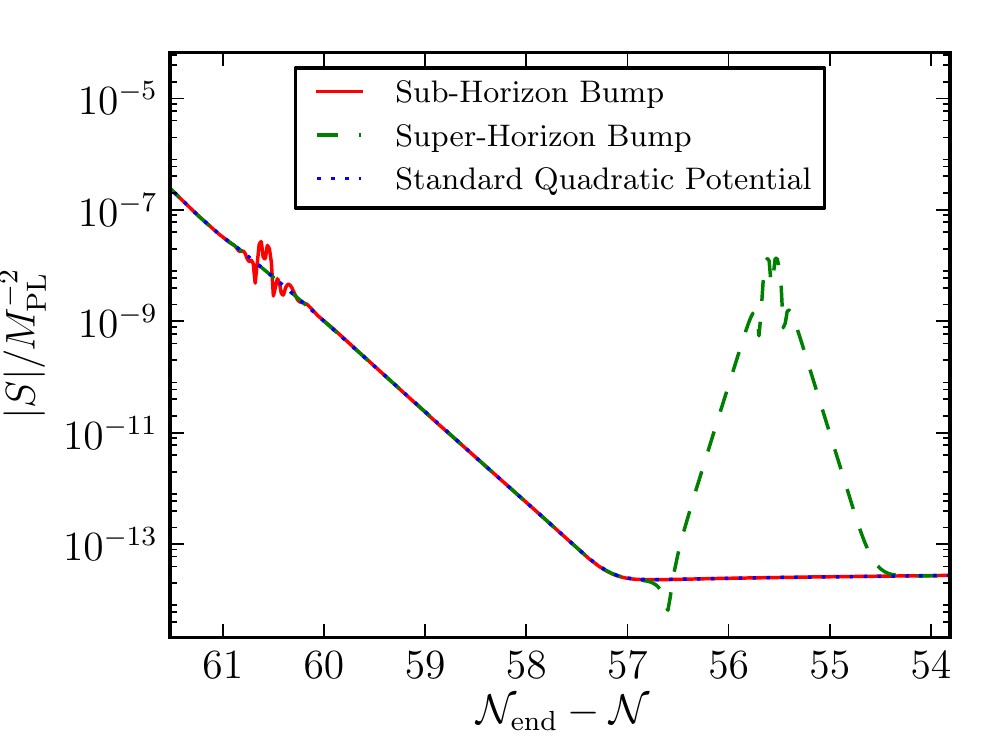}
\label{fig:src-bump-subandsuper-zoom}
}\\
\subfloat[The second order results around the feature.]{
\includegraphics[width=\twofigwidth]{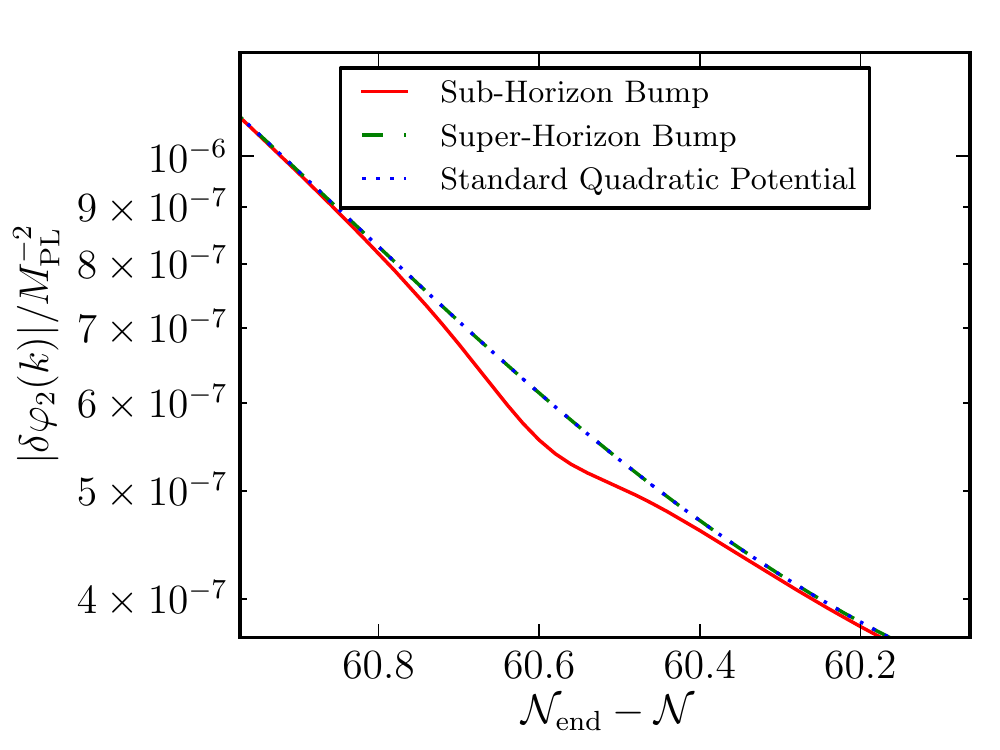}
\label{fig:dp2-bump-subandsuper-zoom}
}
\subfloat[The ratio of the second order results for the sub- and super-horizon bump compared
to the standard quadratic result.]{
\includegraphics[width=\twofigwidth]{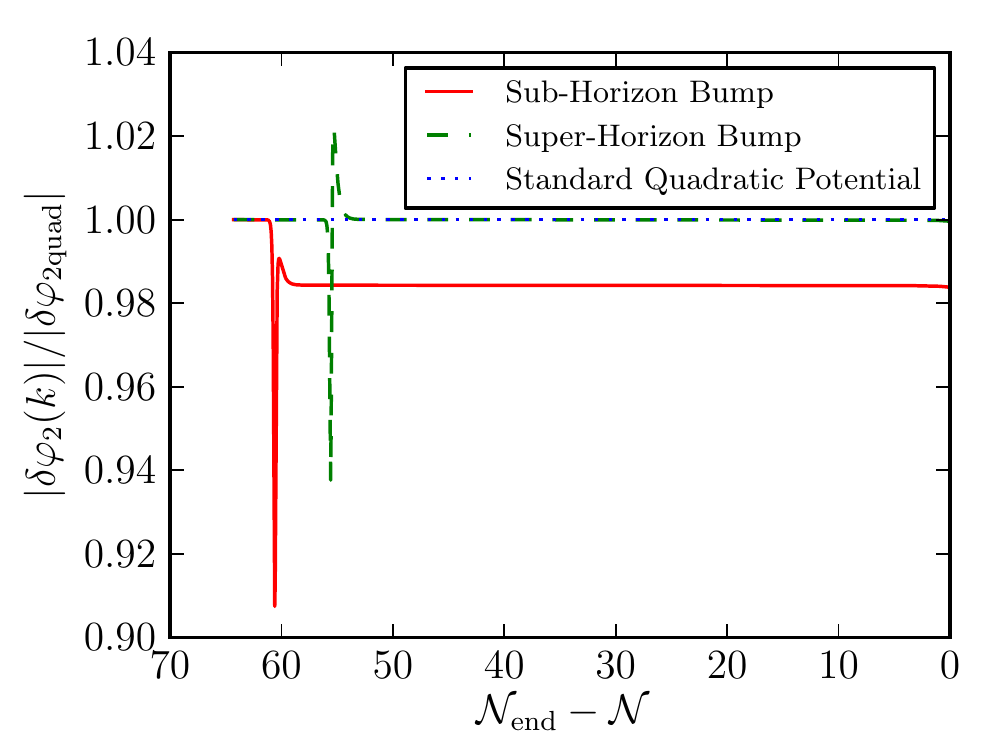}
\label{fig:dp2-bump-subandsuper-ratio}
}
\caption{The first order and source term results for the bump potential when the bump is inside the
horizon when $\vp_{\rm{b}}=15.5\Mpl$ (red solid line) or outside the horizon when
$\vp_{\rm{b}}=14.8\Mpl$ (green
dashed line). The standard quadratic potential results are also shown as the blue dotted line and
all the results are for the scale $\kwmap$.}
\label{fig:dp1-bump-subandsuper-both}
\end{figure}
% % % % % % % % % % % % % % % % % % 
% 
For different $k$ modes the feature occurs when the mode is either
inside or outside the horizon.  When the mode has already crossed the
horizon the result is as in Figure~\ref{fig:dp1-step-both}.  Modes
which are inside the horizon when they encounter the feature in the
potential are not affected in the same way. Of course the location of the bump can
also be changed by varying the $\vp_{\rm{b}}$ parameter
in the potential. In Figure~\ref{fig:dp1-bump-subandsuper-both} the
first order power spectrum and source term results are plotted for a
potential where the bump feature is located inside the horizon and
$\vp_{\rm{b}}=15.5\Mpl$ and compared with the standard quadratic
potential and the normal bump potential where $\vp_{\rm{b}}=14.8\Mpl$.  When
the bump is sub-horizon (the red solid line in
Figure~\ref{fig:dp1-bump-subandsuper-both}) the first order results
are slightly perturbed from the standard result, reaching a slightly
altered magnitude after horizon crossing. This does not happen for the
super-horizon bump (green dashed line) which asymptotes back to the
quadratic result after the feature. 

The source term results also
differ depending on the location of the bump. When the bump is
sub-horizon the change in the magnitude of the source term is much
more suppressed compared to when the bump is outside the horizon. In
the second case the oscillations of nearly all the modes have already
been dampened by the time the feature is encountered. The convolution
integral over the modes is then much more affected by the change in
the first order perturbations around the feature. In contrast when the
bump is sub-horizon, at least for the $\kwmap$ scale, most of the
other scales considered in the integral are still oscillating and the
net effect is a small change in the magnitude of the source term. At
second order the results are similar to first order. The
sub-horizon bump slightly changes the magnitude away from the
quadratic result, in this case reducing it. This change is kept beyond
the horizon in contrast to the super-horizon case where the result
asymptotes smoothly back to the quadratic one beyond the feature. 
Figure~\ref{fig:dp2-bump-subandsuper-ratio} shows the ratio of the 
sub- and super-horizon bump second order results to the standard quadratic 
result. Following the sub-horizon bump the difference between the results
is of the order of 2\% whereas for the super-horizon bump the results are indistinguishable
from the quadratic results after the feature region.
\ \\

% % % % % % % % % % % % % 
In this section we have outlined the main results of our new numerical calculation.
We have shown that the source term results using the full non-slow-roll equation 
\eqref{eq:fullsrc-res} are more damped than the corresponding slow-roll results.
We have demonstrated the new code by using feature potentials with a step and 
a bump added to the standard quadratic potential. Depending on the position of
the feature and the form of the addition to the potential the effects of the
feature can be seen in the source term and second order results beyond the
feature itself. Whether the feature is sub- or super-horizon for a particular
mode also affects the subsequent evolution.

\section{Discussion and Conclusion}
\label{sec:disc}

We describe in this paper the numerical solution of the full
Klein-Gordon equation for a single scalar field at second order in
cosmological perturbation theory. We use gauge-invariant variables in
the flat gauge without imposing the slow-roll approximation or using
the large scale limit. This is an extension of previous work, that
relied on the slow-roll approximation to calculate the source term
\cite{hustonmalik}. We have made this extended code
publicly available (it can be downloaded from the website \cite{pyflation}). 
To validate and test the code, we studied several single
field potentials, including step and bump extensions of the simple
chaotic inflation potential, which violate one of the slow-roll
conditions.

We have shown that feature potentials affect the source and second
order results and at least for the step potential these effects are
apparent throughout the evolution of the modes and not just in a
narrow region around the feature. This is consistent with the higher
order derivatives of the potentials being non-zero, or at least
having different values when computed without a feature.
Is this a problem for feature potentials? At one level the feature
potentials we have used are theoretical toy models which are not
motivated by high energy physics. At first order in perturbation
theory the $\tanh$ and $\cosh$ additions to the quadratic potential
work to localise the deviation to a specific region. This is clearly
not the case at second order for the step potential. 
To replicate this behaviour at second order, a restricted version of
the step or bump potentials could be used.  However, one main benefit
of using $\tanh$ or $\cosh$ is that they smoothly asymptote to the original
potential away from the feature. Any attempted patching together of
featureless potential and feature in a specific region would have to
be carefully constructed to avoid discontinuities in the derivatives
of the potential. In contrast the results for the bump potential are
only affected in a narrow region around the feature.
It is interesting to note, that a step-potential of the form
\eqref{V_bump} has recently been used to study the generation of
magnetic fields in the early universe \cite{Durrer:2010mq}.

Somewhat surprising was the effect the small \emph{sub-horizon} bump
had on the evolution of the second order field fluctuation $\dvp2$ on
large scales, see Figure~\ref{fig:dp1-bump-subandsuper-both}. Sub-horizon effects
like this can only be studied by codes like the one described here,
but would not be {calculable} by super-horizon codes, such as the
recently proposed one in Refs.~\cite{Mulryne:2009kh,Mulryne:2010rp}. However,
it is not clear yet whether this change in $\dvp2$ would translate
into observational consequences, such as a change in $\fnl$. We are
cautiously optimistic that it might have an observable effect, but
postpone a detailed calculation to future work.
In this future work we plan to study more complicated and also more
interesting potentials than in Section \ref{sec:results}, and also
hope to extend the code to allow the evolution of more than one field.
In addition, to our knowledge there is currently no
expression that exists in closed form for $\zeta_2$ or $\mathcal{R}_2$ which is valid on all scales
without imposing slow-roll. Constructing such an expression should be possible and would allow for
direct calculation of the evolution of $\zeta_2$ and we look forward to making progress on this
issue. 
A recent work discussing the numerical calculation of $\fnl$ used
lattice field theory simulations \cite{Barnaby:2010ke}. Whether in future
this method will prove more efficient than the one discussed here will
depend on detailed comparative analyses which are beyond the scope of
the present article. 
Other recent work by Takamizu \etal \cite{Takamizu:2010xy} matched 
a perturbative solution for the curvature of a single field model inside 
the horizon to a non-linear solution outside the
horizon which uses a gradient expansion. The authors claim this goes beyond the
limitations of the $\delta N$ formalism and can also deal with
temporary violations of slow-roll. It would be interesting in future work
to compare the results of our approach with those of Takamizu \etal especially in the region 
where the matching of the solutions takes place. 
Gong, Noh and Hwang have also looked at higher order perturbations \cite{Gong:2010yk}. Their work
concentrates on evolving the convolution kernels of the curvature perturbation but only for large
scales with slow roll parameters equated to zero.
The ``Generalized Slow Roll'' approximation can also be used to investigate models which
break slow roll transiently around a feature \cite{Stewart:2001cd, Dvorkin:2009ne, Adshead:2011bw}.
Adshead \etal \cite{Adshead:2011bw} have
demonstrated calculations of the bispectrum of a single field model and observe good 
agreement with analytic results. However, this approach is limited to superhorizon scales only
and can only be applied when there is a single degree of freedom.  
Another natural application and extension of our code is to apply it
to other problems that have evolution equations of a similar form. As
pointed out in the introduction, two immediate applications might be
the numerical study of the generation of gravitational waves at second
order \cite{Ananda:2006af} (see \Rref{Baumann:2007zm} for earlier numerical
work) and the generation of vorticity \cite{Christopherson:2009bt}.
In both cases the source term of the second order quantities is
given by a convolution integral quadratic in first order quantities.

Finally, the numerical calculation of the convolution integral requires cutoffs to 
be implemented at both large and small scales. We have explained the 
cutoffs we use explicitly in Refs.~\cite{hustonmalik,Huston:2010by}
and refer the interested reader to the discussions contained therein.
It will be interesting to implement in future
work different cut-off schemes, as outlined for example in
Refs.~\cite{Marozzi:2011da,Seery:2010kh}. 
Our current choice of cut-offs is pragmatic, however there may be
physical considerations for adopting a different cut-off scheme, which
would impact on the higher order perturbations and therefore affect the
non-linear observables.

%%%%%%%%%%%%%%%%%%%%%%%%%%%%%%%%%%%%%%%%%%%%%%%%%%%%%%%%%%%%%%%%%%%%%%%

\section*{Acknowledgements}
We would like to thank Kazuya Koyama and Mairi Sakellariadou for their
help, and David Mulryne for useful discussions. IH is supported by
the STFC under Grant ST/G002150/1. KAM is supported in part by the
STFC under Grants ST/G002150/1 and ST/H002855/1.

%%%%%%%%%%%%%%%%%%%%%%%%%%%%%%%%%%%%%%%%%%%%%%%%%%%%%%%%%%%%%%%%%%%%%%%%

\appendix
\section{Appendix}

%%%%%%%%%%%%%%%%%%%%%%%%%%%%%%%%%%%%%
% 
\label{sec:apxsoperts}
%%%%%%%%%%%%%%%%%%%%%%%%%%%%%%%%%%%%%
The methods adopted for the study of first order perturbations can be extended at second
order to find gauge invariant quantities. Recall that scalar quantities such as the inflaton
field, $\varphi$, can be split into an homogeneous background, $\vp_0$, and
inhomogeneous perturbations. Up to second order $\vp$ becomes
\begin{equation}
 \varphi(\eta, x^\mu) = \vp_0(\eta) + \dvp1(\eta, x^i) + \frac{1}{2}\dvp2(\eta, x^i)
\,.
\end{equation}

The metric tensor $g_{\mu\nu}$ must also be perturbed up to second order. Here we consider only the
scalar metric perturbations \cite{Malik:2008im}:
\begin{align}
\label{eq:metric1-num}
g_{00}&= -a^2\left(1+2\phi_1+\phi_2\right) \,, \nonumber\\
g_{0i}&= a^2\left(B_1+\frac{1}{2}B_2\right)_{,i}\,, \nonumber\\
g_{ij}&= a^2\left[\left(1-2\psi_1-\psi_2\right)\delta_{ij}
+2E_{1,ij}+E_{2,ij}\right]\,,
\end{align}
where $\delta_{ij}$ is the flat background metric, $\phi_1$ and $\phi_2$ are the
lapse functions at first and second order, $\psi_1$ and $\psi_2$ are the curvature
perturbations, and $B_1$, $B_2$, $E_1$ and $E_2$ are the
scalar perturbations describing the shear.
As well as the first order transformation vector, there
is a second order transformation vector and they are both given by
\begin{equation}
 \label{eq:xi2defn-perts}
\xi_1^\mu = (\alpha_1, \beta_{1,}^{~~i})\,,\quad \xi_2^\mu = (\alpha_2, \beta_{2,}^{~~i}) \,,
\end{equation}
where the spatial vector part of the transformation has been ignored. 

The transformation of a second order scalar quantity (such as $\dvp2$) is given by
\cite{Malik:2003mv,Malik:2008im}:
\begin{equation}
\label{eq:dvp2transform-perts}
 \wt{\dvp2} = \dvp2 + \vp_0'\alpha_2 + \alpha_1\left(\vp_0'' \alpha_1 + \vp_0'
\alpha_1' + 2\dvp1'\right) + \left(2\dvp1 + \vp_0'\alpha_1\right)_{,i}
\beta_{1,}^{~~i} \,,
\end{equation}
where a tilde ($\wt{~}$) denotes a transformed quantity. 
The metric curvature perturbation transformation at first order is straightforward,
$\wt{\psi_1} = \psi_1 -\H \alpha_1$, but at second order it becomes more complicated
\cite{Malik:2008yp,Malik:2008im}:
\begin{equation}
 \label{eq:psi2transform-perts}
\wt{\psi_2} = \psi_2 -\H\alpha_2 -\frac{1}{4}\mathcal{X}^k_{~k} +
 \frac{1}{4}\nabla^{-2} \mathcal{X}^{ij}_{~~ij}\,,
\end{equation}
where $\mathcal{X}_{ij}$ is given by
\begin{align}
 \label{Xijdef}
\mathcal{X}_{ij} \equiv 
&2\Big[\left(\H^2+\frac{a''}{a}\right)\alpha_1^2
+\H\left(\alpha_1\alpha_1'+\alpha_{1,k}\xi_{1}^{~k}
\right)\Big] \delta_{ij}\nonumber\\
&
+4\Big[\alpha_1\left(C_{1ij}'+2\H C_{1ij}\right)
+C_{1ij,k}\xi_{1}^{~k}+C_{1ik}\xi_{1~~,j}^{~k}
+C_{1kj}\xi_{1~~,i}^{~k}\Big] \nonumber\\
&+2\left(B_{1i}\alpha_{1,j}+B_{1j}\alpha_{1,i}\right)
+4\H\alpha_1\left( \xi_{1i,j}+\xi_{1j,i}\right)
-2\alpha_{1,i}\alpha_{1,j}+2\xi_{1k,i}\xi_{1~~,j}^{~k} \nonumber \\
&+\alpha_1\left( \xi_{1i,j}'+\xi_{1j,i}' \right)
+\left(\xi_{1i,jk}+\xi_{1j,ik}\right)\xi_{1}^{~k}
+\xi_{1i,k}\xi_{1~~,j}^{~k}+\xi_{1j,k}\xi_{1~~,i}^{~k} \nonumber \\
&+\xi_{1i}'\alpha_{1,j}+\xi_{1j}'\alpha_{1,i}
\,,
\end{align}
and $B_{1i}$ and $C_{1ij}$ are defined as
\begin{equation}
\label{eq:svt-intro}
  B_{1i} = B_{1,i} \,, \quad 
  C_{1ij} = -\psi_1\delta_{ij} + E_{1,ij} \,.
\end{equation}
Working in the uniform curvature gauge, where spatial 3-hypersurfaces are flat, implies
that
\begin{equation}
 \label{eq:gauge-num}
\wt\psi_1=\wt\psi_2=\wt E_1=\wt E_2=0 \,.
\end{equation}
These relations can be used at first and then at second order to define gauge
invariant variables. The first order
transformation variables in the flat gauge satisfy $\alpha_1 = \psi_1/\H$ and
$\beta_1 = -E_1$. At second order, for the transformation of scalar quantities, as in
\eq{eq:dvp2transform-perts}, we require only $\alpha_2$. This is found by using
\eq{eq:psi2transform-perts} to have the form
\begin{equation}
 \label{eq:alpha2-perts}
\alpha_2=\frac{\psi_2}{\H}+\frac{1}{4\H}\left[
\nabla^{-2}\X^{ij}_{~~,ij}-\X^k_{ k}\right]\,,
\end{equation}
where the first order gauge variables have been substituted into $\X_{ij}$.

The Sasaki-Mukhanov variable, \iec the field perturbation on uniform curvature
hypersurfaces \cite{Sasaki:1986hm,Mukhanov:1988jd}, is given at first order by
\begin{equation}
\label{eq:Q1I-num}
\wt{\dvp1}=\dvp1+\frac{\vp_{0}'}{\H}\psi_1\,.
\end{equation}
At second order the Sasaki-Mukhanov variable becomes more complicated
\cite{Malik:2005cy,Malik:2003mv}:
\begin{align}
\label{eq:Q2I-num}
\wt{\dvp2} = &\dvp2
+\frac{\vp_0'}{\H}\psi_2
+\frac{\vp_0'}{4\H}\left(
\nabla^{-2}\X^{ij}_{~~,ij}-\X^k_{k}\right) \nonumber\\
&+\frac{\psi_1}{\H^2}\Big[\vp_0'' {\psi_1}
+\vp_0'\left(\psi_1'-\frac{\H'}{\H}\psi_1\right)+2\H\delta\vp_1'\Big]
+\left(2\dvp1+\frac{\vp_0'}{\H}\psi_1\right)_{,k}\xi_{1\mathrm{flat}}^k \,,
\end{align}
where $\xi_{1\mathrm{flat}} = -E_{1,i}$. From now on we will
drop the tildes and only refer to variables calculated in the flat gauge.
The potential of the scalar field can also be separated into homogeneous and
perturbed sectors:
\begin{align}
 V(\varphi) &= \U + \delta V_{1} + \frac{1}{2}\delta V _{2}\,,\quad \\
 \delta V_{1} &= \Uphi \dvp1 \,,\quad \\
 \delta V_{2} &= \Upp \dvp1^2 + \Uphi\dvp2 \,.
\end{align}
The Klein-Gordon equations are found by requiring the perturbed energy-momentum tensor $T_{\mu\nu}$
to obey the energy conservation equation $\nabla_\mu T^{\mu\nu}=0$ (see for example
\Rref{Malik:2005cy}). For the
background field, $\vp_0$, the Klein-Gordon equation is 
\begin{equation}
\label{eq:KGback-num}
\vp_{0}''+2\H\vp_{0}'+a^2 \Uphi = 0\,.   
\end{equation}
The first order equation is
\begin{equation}
\label{eq:KGflatsingle-num}
\dvp1''+2\H\dvp1'+2a^2 \Uphi \phi_1
-\nabla^2\dvp1-\vp_{0}'\nabla^2 B_1
-\vp_{0}'\phi'_1 + a^2 \Upp \dvp1
=0\,,
\end{equation}
and the second order one is given by
\begin{align}
\label{eq:KG2flatsingle-num}
\dvp2'' &+ 2\H\dvp2'-\nabla^2\dvp2+a^2 \Upp \dvp2
+ a^2 \Uppp (\dvp1)^2 +2a^2 \Uphi \phi_2
-\vp_{0}'\left(\nabla^2 B_2+\phi_2'\right)\nonumber\\
&+ 4\vp_{0}' B_{1,k}\phi_{1,}^{~k}
+2\left(2\H\vp_{0}'+a^2 \Uphi\right) B_{1,k}B_{1,}^{~k}
+4\phi_1\left(a^2 \Upp \dvp1-\nabla^2\dvp1\right) \nonumber\\
&+ 4\vp_{0}'\phi_1\phi_1'
-2\dvp1'\left(\nabla^2 B_1+\phi_1'\right)-4{\dvp1'}_{,k}B_{1,}^{~k} \nonumber \\
&= 0\,,
\end{align}
where all the variables are now in the flat gauge.

In order to write the Klein-Gordon equations in closed form, the Einstein field
equations are also required at first and second order.
These are not reproduced here, but are presented for example in Section~II~B of
\Rref{Malik:2006ir}. 

% Bibliography
% \bibliography{central-store}

\begin{thebibliography}{10}

\bibitem{wmapwebsite}
{NASA/WMAP Team}, ``{WMAP} mission website.'' \url{http://wmap.gsfc.nasa.gov}.

\bibitem{planckwebsite}
{ESA/Planck Team}, ``{Planck} mission website.''
  \url{http://www.esa.int/planck}.

\bibitem{book:liddle}
A.~R. Liddle and D.~H. Lyth, {\em Cosmological Inflation and Large-Scale
  Structure}.
\newblock Cambridge University Press, April, 2000.

\bibitem{Leach:2001zf}
S.~M. Leach, M.~Sasaki, D.~Wands, and A.~R. Liddle, ``{Enhancement of
  superhorizon scale inflationary curvature perturbations},''
  \href{http://dx.doi.org/10.1103/PhysRevD.64.023512}{{\em Phys. Rev.} {\bf
  D64} (2001)  023512},
\href{http://arxiv.org/abs/astro-ph/0101406}{{\tt arXiv:astro-ph/0101406}}.
%%CITATION = ASTRO-PH/0101406;%%.

\bibitem{Malik:2005cy}
K.~A. Malik, ``{Gauge-invariant perturbations at second order: Multiple scalar
  fields on large scales},'' {\em JCAP} {\bf 0511} (2005)  005,
\href{http://arxiv.org/abs/astro-ph/0506532}{{\tt arXiv:astro-ph/0506532}}.
%%CITATION = ASTRO-PH/0506532;%%.

\bibitem{Malik:2007nd}
K.~A. Malik, D.~Seery, and K.~N. Ananda, ``{Different approaches to the second
  order Klein-Gordon equation},''
  \href{http://dx.doi.org/10.1088/0264-9381/25/17/175008}{{\em Class. Quant.
  Grav.} {\bf 25} (2008)  175008},
\href{http://arxiv.org/abs/0712.1787}{{\tt arXiv:0712.1787 [astro-ph]}}.
%%CITATION = 0712.1787;%%.

\bibitem{Malik:2006ir}
K.~A. Malik, ``{A not so short note on the Klein-Gordon equation at second
  order},'' {\em JCAP} {\bf 0703} (2007)  004,
\href{http://arxiv.org/abs/astro-ph/0610864}{{\tt arXiv:astro-ph/0610864}}.
%%CITATION = ASTRO-PH/0610864;%%.

\bibitem{hustonmalik}
I.~Huston and K.~A. Malik, ``{Numerical calculation of second order
  perturbations},'' \href{http://dx.doi.org/10.1088/1475-7516/2009/09/019}{{\em
  JCAP} {\bf 0909} (2009)  019},
\href{http://arxiv.org/abs/0907.2917}{{\tt arXiv:0907.2917 [astro-ph.CO]}}.
%%CITATION = 0907.2917;%%.

\bibitem{Adams:2001vc}
J.~A. Adams, B.~Cresswell, and R.~Easther, ``{Inflationary perturbations from a
  potential with a step},''
  \href{http://dx.doi.org/10.1103/PhysRevD.64.123514}{{\em Phys. Rev.} {\bf
  D64} (2001)  123514},
\href{http://arxiv.org/abs/astro-ph/0102236}{{\tt arXiv:astro-ph/0102236}}.
%%CITATION = ASTRO-PH/0102236;%%.

\bibitem{Stewart:2001cd}
E.~D. Stewart, ``{The spectrum of density perturbations produced during
  inflation to leading order in a general slow-roll approximation},''
  \href{http://dx.doi.org/10.1103/PhysRevD.65.103508}{{\em Phys. Rev.} {\bf
  D65} (2002)  103508},
\href{http://arxiv.org/abs/astro-ph/0110322}{{\tt arXiv:astro-ph/0110322}}.
%%CITATION = ASTRO-PH/0110322;%%.

\bibitem{Choe:2004zg}
J.~Choe, J.-O. Gong, and E.~D. Stewart, ``{Second order general slow-roll power
  spectrum},'' \href{http://dx.doi.org/10.1088/1475-7516/2004/07/012}{{\em
  JCAP} {\bf 0407} (2004)  012},
\href{http://arxiv.org/abs/hep-ph/0405155}{{\tt arXiv:hep-ph/0405155}}.
%%CITATION = HEP-PH/0405155;%%.

\bibitem{Joy:2008qd}
M.~Joy, A.~Shafieloo, V.~Sahni, and A.~A. Starobinsky, ``{Is a step in the
  primordial spectral index favored by CMB data ?},''
  \href{http://dx.doi.org/10.1088/1475-7516/2009/06/028}{{\em JCAP} {\bf 0906}
  (2009)  028},
\href{http://arxiv.org/abs/0807.3334}{{\tt arXiv:0807.3334 [astro-ph]}}.
%%CITATION = 0807.3334;%%.

\bibitem{Hamann:2009bz}
J.~Hamann, A.~Shafieloo, and T.~Souradeep, ``{Features in the primordial power
  spectrum? A frequentist analysis},''
  \href{http://dx.doi.org/10.1088/1475-7516/2010/04/010}{{\em JCAP} {\bf 1004}
  (2010)  010},
\href{http://arxiv.org/abs/0912.2728}{{\tt arXiv:0912.2728 [astro-ph.CO]}}.
%%CITATION = 0912.2728;%%.

\bibitem{Hazra:2010ve}
D.~K. Hazra, M.~Aich, R.~K. Jain, L.~Sriramkumar, and T.~Souradeep,
  ``{Primordial features due to a step in the inflaton potential},''
  \href{http://dx.doi.org/10.1088/1475-7516/2010/10/008}{{\em JCAP} {\bf 1010}
  (2010)  008},
\href{http://arxiv.org/abs/1005.2175}{{\tt arXiv:1005.2175 [astro-ph.CO]}}.
%%CITATION = 1005.2175;%%.

\bibitem{Chen:2008wn}
X.~Chen, R.~Easther, and E.~A. Lim, ``{Generation and Characterization of Large
  Non-Gaussianities in Single Field Inflation},''
  \href{http://dx.doi.org/10.1088/1475-7516/2008/04/010}{{\em JCAP} {\bf 0804}
  (2008)  010},
\href{http://arxiv.org/abs/0801.3295}{{\tt arXiv:0801.3295 [astro-ph]}}.
%%CITATION = 0801.3295;%%.

\bibitem{Chen:2006xjb}
X.~Chen, R.~Easther, and E.~A. Lim, ``{Large non-Gaussianities in single field
  inflation},'' {\em JCAP} {\bf 0706} (2007)  023,
\href{http://arxiv.org/abs/astro-ph/0611645}{{\tt arXiv:astro-ph/0611645}}.
%%CITATION = ASTRO-PH/0611645;%%.

\bibitem{Starobinsky:1982ee}
A.~A. Starobinsky, ``{Dynamics of Phase Transition in the New Inflationary
  Universe Scenario and Generation of Perturbations},''
\href{http://dx.doi.org/10.1016/0370-2693(82)90541-X}{{\em Phys. Lett.} {\bf
  B117} (1982)  175--178}.
%%CITATION = PHLTA,B117,175;%%.

\bibitem{Starobinsky:1986fxa}
A.~A. Starobinsky, ``{Multicomponent de Sitter (Inflationary) Stages and the
  Generation of Perturbations},''
{\em JETP Lett.} {\bf 42} (1985)  152--155.
%%CITATION = JTPLA,42,152;%%.

\bibitem{Salopek:1990jq}
D.~S. Salopek and J.~R. Bond, ``{Nonlinear evolution of long wavelength metric
  fluctuations in inflationary models},''
\href{http://dx.doi.org/10.1103/PhysRevD.42.3936}{{\em Phys. Rev.} {\bf D42}
  (1990)  3936--3962}.
%%CITATION = PHRVA,D42,3936;%%.

\bibitem{Sasaki:1995aw}
M.~Sasaki and E.~D. Stewart, ``{A General analytic formula for the spectral
  index of the density perturbations produced during inflation},''
  \href{http://dx.doi.org/10.1143/PTP.95.71}{{\em Prog. Theor. Phys.} {\bf 95}
  (1996)  71--78},
\href{http://arxiv.org/abs/astro-ph/9507001}{{\tt arXiv:astro-ph/9507001}}.
%%CITATION = ASTRO-PH/9507001;%%.

\bibitem{Lyth:2004gb}
D.~H. Lyth, K.~A. Malik, and M.~Sasaki, ``{A general proof of the conservation
  of the curvature perturbation},''
  \href{http://dx.doi.org/10.1088/1475-7516/2005/05/004}{{\em JCAP} {\bf 0505}
  (2005)  004},
\href{http://arxiv.org/abs/astro-ph/0411220}{{\tt arXiv:astro-ph/0411220}}.
%%CITATION = ASTRO-PH/0411220;%%.

\bibitem{Ananda:2006af}
K.~N. Ananda, C.~Clarkson, and D.~Wands, ``{The cosmological gravitational wave
  background from primordial density perturbations},''
  \href{http://dx.doi.org/10.1103/PhysRevD.75.123518}{{\em Phys. Rev.} {\bf
  D75} (2007)  123518},
\href{http://arxiv.org/abs/gr-qc/0612013}{{\tt arXiv:gr-qc/0612013}}.
%%CITATION = GR-QC/0612013;%%.

\bibitem{Christopherson:2009bt}
A.~J. Christopherson, K.~A. Malik, and D.~R. Matravers, ``{Vorticity generation
  at second order},'' \href{http://dx.doi.org/10.1103/PhysRevD.79.123523}{{\em
  Phys. Rev.} {\bf D79} (2009)  123523},
\href{http://arxiv.org/abs/0904.0940}{{\tt arXiv:0904.0940 [astro-ph.CO]}}.
%%CITATION = 0904.0940;%%.

\bibitem{pyflation}
I.~Huston, ``{Pyflation}: Cosmological perturbations for {Python}.''
  \url{http://pyflation.ianhuston.net}, 2011--.

\bibitem{Malik:2008yp}
K.~A. Malik and D.~R. Matravers, ``{A Concise Introduction to Perturbation
  Theory in Cosmology},''
  \href{http://dx.doi.org/10.1088/0264-9381/25/19/193001}{{\em Class. Quant.
  Grav.} {\bf 25} (2008)  193001},
\href{http://arxiv.org/abs/0804.3276}{{\tt arXiv:0804.3276 [astro-ph]}}.
%%CITATION = 0804.3276;%%.

\bibitem{Malik:2008im}
K.~A. Malik and D.~Wands, ``{Cosmological perturbations},''
  \href{http://dx.doi.org/10.1016/j.physrep.2009.03.001}{{\em Phys. Rept.} {\bf
  475} (2009)  1--51},
\href{http://arxiv.org/abs/0809.4944}{{\tt arXiv:0809.4944 [astro-ph]}}.
%%CITATION = 0809.4944;%%.

\bibitem{Noh:2004bc}
H.~Noh and J.-c. Hwang, ``{Second-order perturbations of the Friedmann world
  model},''
\href{http://dx.doi.org/10.1103/PhysRevD.69.104011}{{\em Phys. Rev.} {\bf D69}
  (2004)  104011}.
%%CITATION = PHRVA,D69,104011;%%.

\bibitem{Vretblad:2005}
A.~Vretblad, {\em {Fourier Analysis and Its Applications (Graduate Texts in
  Mathematics)}}.
\newblock Springer, 2005.

\bibitem{Huston:2010by}
I.~Huston, ``{Constraining Inflationary Scenarios with Braneworld Models and
  Second Order Cosmological Perturbations},''
\href{http://arxiv.org/abs/1006.5321}{{\tt arXiv:1006.5321 [astro-ph.CO]}}.
%%CITATION = 1006.5321;%%.

\bibitem{Salopek:1988qh}
D.~S. Salopek, J.~R. Bond, and J.~M. Bardeen, ``{Designing Density Fluctuation
  Spectra in Inflation},''
\href{http://dx.doi.org/10.1103/PhysRevD.40.1753}{{\em Phys. Rev.} {\bf D40}
  (1989)  1753}.
%%CITATION = PHRVA,D40,1753;%%.

\bibitem{Ringeval:2007am}
C.~Ringeval, ``{The exact numerical treatment of inflationary models},''
  \href{http://dx.doi.org/10.1007/978-3-540-74353-8_7}{{\em Lect. Notes Phys.}
  {\bf 738} (2008)  243--273},
\href{http://arxiv.org/abs/astro-ph/0703486}{{\tt arXiv:astro-ph/0703486}}.
%%CITATION = ASTRO-PH/0703486;%%.

\bibitem{Martin:2006rs}
J.~Martin and C.~Ringeval, ``{Inflation after WMAP3: Confronting the slow-roll
  and exact power spectra to CMB data},'' {\em JCAP} {\bf 0608} (2006)  009,
\href{http://arxiv.org/abs/astro-ph/0605367}{{\tt arXiv:astro-ph/0605367}}.
%%CITATION = ASTRO-PH/0605367;%%.

\bibitem{Durrer:2010mq}
R.~Durrer, L.~Hollenstein, and R.~K. Jain, ``{Can slow roll inflation induce
  relevant helical magnetic fields?},''
  \href{http://arxiv.org/abs/1005.5322}{{\tt arXiv:1005.5322 [astro-ph.CO]}}.

\bibitem{Mulryne:2009kh}
D.~J. Mulryne, D.~Seery, and D.~Wesley, ``{Moment transport equations for
  non-Gaussianity},''
  \href{http://dx.doi.org/10.1088/1475-7516/2010/01/024}{{\em JCAP} {\bf 1001}
  (2010)  024},
\href{http://arxiv.org/abs/0909.2256}{{\tt arXiv:0909.2256 [astro-ph.CO]}}.
%%CITATION = 0909.2256;%%.

\bibitem{Mulryne:2010rp}
D.~J. Mulryne, D.~Seery, and D.~Wesley, ``{Moment transport equations for the
  primordial curvature perturbation},''
\href{http://arxiv.org/abs/1008.3159}{{\tt arXiv:1008.3159 [astro-ph.CO]}}.
%%CITATION = 1008.3159;%%.

\bibitem{Barnaby:2010ke}
N.~Barnaby, ``{On Features and Nongaussianity from Inflationary Particle
  Production},'' \href{http://dx.doi.org/10.1103/PhysRevD.82.106009}{{\em
  Phys.Rev.} {\bf D82} (2010)  106009},
  \href{http://arxiv.org/abs/1006.4615}{{\tt arXiv:1006.4615 [astro-ph.CO]}}.

\bibitem{Takamizu:2010xy}
Y.-i. Takamizu, S.~Mukohyama, M.~Sasaki, and Y.~Tanaka, ``{Non-Gaussianity of
  superhorizon curvature perturbations beyond $\delta$ N formalism},''
  \href{http://dx.doi.org/10.1088/1475-7516/2010/06/019}{{\em JCAP} {\bf 1006}
  (2010)  019},
\href{http://arxiv.org/abs/1004.1870}{{\tt arXiv:1004.1870 [astro-ph.CO]}}.
%%CITATION = 1004.1870;%%.

\bibitem{Gong:2010yk}
J.-O. Gong, H.~Noh, and J.-c. Hwang, ``{Non-linear corrections to inflationary
  power spectrum},''
  \href{http://dx.doi.org/10.1088/1475-7516/2011/04/004}{{\em JCAP} {\bf 1104}
  (2011)  004},
\href{http://arxiv.org/abs/1011.2572}{{\tt arXiv:1011.2572 [astro-ph.CO]}}.
%%CITATION = 1011.2572;%%.

\bibitem{Dvorkin:2009ne}
C.~Dvorkin and W.~Hu, ``{Generalized Slow Roll for Large Power Spectrum
  Features},'' \href{http://dx.doi.org/10.1103/PhysRevD.81.023518}{{\em Phys.
  Rev.} {\bf D81} (2010)  023518},
\href{http://arxiv.org/abs/0910.2237}{{\tt arXiv:0910.2237 [astro-ph.CO]}}.
%%CITATION = 0910.2237;%%.

\bibitem{Adshead:2011bw}
P.~Adshead, W.~Hu, C.~Dvorkin, and H.~V. Peiris, ``{Fast Computation of
  Bispectrum Features with Generalized Slow Roll},''
\href{http://arxiv.org/abs/1102.3435}{{\tt arXiv:1102.3435 [astro-ph.CO]}}.
%%CITATION = 1102.3435;%%.

\bibitem{Baumann:2007zm}
D.~Baumann, P.~J. Steinhardt, K.~Takahashi, and K.~Ichiki, ``{Gravitational
  Wave Spectrum Induced by Primordial Scalar Perturbations},''
  \href{http://dx.doi.org/10.1103/PhysRevD.76.084019}{{\em Phys.Rev.} {\bf D76}
  (2007)  084019}, \href{http://arxiv.org/abs/hep-th/0703290}{{\tt
  arXiv:hep-th/0703290 [hep-th]}}.

\bibitem{Marozzi:2011da}
G.~Marozzi, M.~Rinaldi, and R.~Durrer, ``{On infrared and ultraviolet
  divergences of cosmological perturbations},''
\href{http://arxiv.org/abs/1102.2206}{{\tt arXiv:1102.2206 [astro-ph.CO]}}.
%%CITATION = 1102.2206;%%.

\bibitem{Seery:2010kh}
D.~Seery, ``{Infrared effects in inflationary correlation functions},''
  \href{http://dx.doi.org/10.1088/0264-9381/27/12/124005}{{\em
  Class.Quant.Grav.} {\bf 27} (2010)  124005},
  \href{http://arxiv.org/abs/1005.1649}{{\tt arXiv:1005.1649 [astro-ph.CO]}}.

\bibitem{Malik:2003mv}
K.~A. Malik and D.~Wands, ``{Evolution of second order cosmological
  perturbations},'' {\em Class. Quant. Grav.} {\bf 21} (2004)  L65--L72,
\href{http://arxiv.org/abs/astro-ph/0307055}{{\tt arXiv:astro-ph/0307055}}.
%%CITATION = ASTRO-PH/0307055;%%.

\bibitem{Sasaki:1986hm}
M.~Sasaki, ``{Large Scale Quantum Fluctuations in the Inflationary Universe},''
\href{http://dx.doi.org/10.1143/PTP.76.1036}{{\em Prog. Theor. Phys.} {\bf 76}
  (1986)  1036}.
%%CITATION = PTPKA,76,1036;%%.

\bibitem{Mukhanov:1988jd}
V.~F. Mukhanov, ``{Quantum Theory of Gauge Invariant Cosmological
  Perturbations},''
{\em Sov. Phys. JETP} {\bf 67} (1988)  1297--1302.
%%CITATION = SPHJA,67,1297;%%.

\end{thebibliography}
% \bibliographystyle{utphys}
% \bibliographystyle{plain}

\providecommand{\href}[2]{#2}\begingroup\raggedright\endgroup

\end{document}